\documentclass[prd,preprint,superscriptaddress,amsmath,amssymb]{revtex4}

\usepackage{graphicx,color,slashed}

\begin{document}

{\begin{flushright}{KIAS-P18083}
\end{flushright}}

\title{\bf  $\epsilon_K$ and $\epsilon'/\epsilon$ in a diquark model}

\author{Chuan-Hung Chen}
\email{physchen@mail.ncku.edu.tw}
\affiliation{Department of Physics, National Cheng-Kung University, Tainan 70101, Taiwan}

\author{Takaaki Nomura}
\email{nomura@kias.re.kr}
\affiliation{School of Physics, KIAS, Seoul 02455, Korea}

\date{\today}

\begin{abstract}
Based on the calculations  using the  lattice QCD by the RBC-UKQCD collaboration and a large $N_c$ dual QCD,  the resulted $\epsilon'/\epsilon$, which is less than  the experimental data by more than a $2\sigma$ in the standard model (SM), suggests the necessity of a new physics effect. In order to complement the insufficient $\epsilon'/\epsilon$, we study the extension of the SM with a colored scalar in a diquark model. In addition to the pure diquark box diagrams, it is found that the box diagrams with one $W$-boson and one diquark, ignored in the literature, play an important role in the $\Delta S=2$ process. The mass difference between $K_L$ and $K_S$ in the diquark model is well below the current data, whereas the Kaon indirect CP violation $\epsilon_K$ gives a strict constraint on the new parameters. Three mechanisms are classified in the study of $\epsilon'/\epsilon$.  They include a tree-level diagram, QCD and electroweak (EW) penguins, and chromomagnetic  operators (CMOs).   Taking the Kobayashi-Maskawa phase as the unique CP source, we analyze each contribution of the three mechanisms in detail and conclude that with the exception of QCD and EW penguins, the tree and CMO effects can singly  enhance $\epsilon'/\epsilon$ to be of ${\cal O}(10^{-3})$, depending on the values of free parameters, when the bound from $\epsilon_K$ is satisfied. 

\end{abstract}

\maketitle

\section{Introduction}

It is known that the measured CP violation in $K$ and $B$ meson decays can be attributed to the unique CP phase of the Cabibbo-Kobayashi-Maskawa (CKM) matrix~\cite{Cabibbo:1963yz,Kobayashi:1973fv} in the standard model (SM). However, it is a long-standing challenge to theoretically predict the Kaon direct CP violation $\epsilon'/\epsilon$ in the SM. Now, the progress in predicting $\epsilon'/\epsilon$ has taken one step forward based on two results: one is from  lattice QCD calculations and the other is a QCD theory-based approach. 

Firstly, the RBC-UKQCD collaboration recently  reported  the surprising lattice QCD results on the matrix elements of $K\to \pi \pi$ and $\epsilon'/\epsilon$~\cite{Boyle:2012ys,Blum:2011ng,Blum:2012uk,Blum:2015ywa,Bai:2015nea}, where the the electroweak (EW) penguin  contribution to $\epsilon'/\epsilon$ and the Kaon direct CP violation are, respectively, shown as~\cite{Blum:2015ywa,Bai:2015nea}:
 \begin{equation}
Re(\epsilon'/\epsilon)_{\rm EWP}= -(6.6 \pm 1.0) \times 10^{-4}\,, \quad  Re(\epsilon'/\epsilon) = 1.38(5.15)(4.59) \times 10^{-4}\,; \label{eq:LQCD}
 \end{equation}
however, the experimental average measured by the  NA48~\cite{Batley:2002gn} and KTeV~\cite{AlaviHarati:2002ye,Abouzaid:2010ny}  is  $Re(\epsilon'/\epsilon)=(16.6 \pm 2.3)\times 10^{-4}$. As a result,   the lattice calculations indicate that the SM prediction is  $2.1\sigma$ below the experimental value. 

Using a large $N_c$ dual QCD (DQCD) approach~\cite{Buras:2015xba, Buras:2015yba},  which was developed by~\cite{Buras:1985yx,Bardeen:1986vp,Bardeen:1986uz,Bardeen:1986vz,Bardeen:1987vg}, the  calculations of $Re(\epsilon'/\epsilon)$ in the QCD based approach support the RBC-UKQCD results, and the results are given as:
 \begin{align}
 Re(\epsilon'/\epsilon)_{\rm SM} &= \left\{
\begin{array}{c}
  (8.6 \pm 3.2)\times 10^{-4}\,, ~~\text(B^{(1/2)}_6=B^{(3/2)}_8=1)\,,       \\
  (6.0 \pm 2.4) \times 10^{-4}\,, ~~ \text(B^{(1/2)}_6=B^{(3/2)}_8=0.76)\,, 
\end{array}
\right.
 \end{align}
where $B^{(1/2)}_6$ and $B^{(3/2)}_8$ denote the non-perturbative parameters of the gluon ($Q_6$) and EW ($Q_8$) penguin operators, respectively.  Regardless of  what the correct values of $B^{1/2}_6$ and $B^{3/2}_8$ are,  the predicted $Re(\epsilon'/\epsilon)_{\rm SM}$ also is over $2\sigma$ below the data. Although the uncertainty of $B^{(1/2)}_6$ is still large, it is found that both approaches obtain consistent values in $B^{(1/2)}_6$ and $B^{(3/2)}_8$ as~\cite{Buras:2015xba}:
 \begin{align}
& B^{(1/2)}_6 (m_c) = 0.57\pm 0.19  \,, \ B^{(3/2)}_8(m_c)=0.76 \pm 0.05 ~~  \text{(RBC-UKQCD)}\,, \nonumber \\
&  B^{(1/2)}_6 \leq B^{(3/2)} < 1 \,, \ B^{(3/2)}_8(m_c)=0.80 \pm 0.1\,. ~~~~~~~~~~~ \text{(large $N_c$)}\,.
 \end{align}
 If the RBC-UKQCD results of  $B^{(1/2)}_6 (m_c) = 0.57\pm 0.19$ and   $B^{(3/2)}_8(m_c)=0.76 \pm 0.05$ are used, the Kaon direct CP violation becomes~\cite{Buras:2015yba}:
 \begin{equation}
 Re(\epsilon'/\epsilon)_{\rm SM} =  (1.9 \pm 4.5)\times 10^{-4}\,,
  %Re(\epsilon'/\epsilon)_{\rm SM} &=  (8.6 \pm 3.2)\times 10^{-4}\,, ~~\text(B^{(1/2)}_6=B^{(3/2)}_8=1)\,, \\
 %
%  Re(\epsilon'/\epsilon)_{\rm SM} & =  (6.0 \pm 2.4) \times 10^{-4}\,, ~~ \text(B^{(1/2)}_6=B^{(3/2)}_8=0.76)\,.
 \end{equation}
where the DQCD's value is even closer to the RBC-UKQCD result shown in Eq.~(\ref{eq:LQCD}). Moreover, using the lattice QCD results, the authors in~\cite{Kitahara:2016nld} also obtained a consistent result with $Re(\epsilon'/\epsilon)=(1.06\pm5.07)\times 10^{-4}$ at the next-leading order (NLO) corrections. 

Since  the DQCD result arises from the short-distance (SD) four-fermion operators,  it is of interest to find other mechanisms that can complement the  insufficient $\epsilon'/\epsilon$, in the SM, such as the  long-distance (LD) final state interactions (FSIs).  However, the conclusion of the LD contribution is still uncertain, where the authors in~\cite{Buras:2016fys} obtained a negative conclusion, but the authors in~\cite{Gisbert:2017vvj} obtained $Re(\epsilon'/\epsilon)=(15\pm7) \times 10^{-4}$ when the SD and LD effects were included. On the other hand, in spite of the  large uncertainty of the current lattice calculations, if we take the RBC-UKQCD's central value as the tendency of the SM,  the alternative source to enhance $\epsilon'/\epsilon$ can be from a new physics effect~\cite{Buras:2015qea,Buras:2015yca,Buras:2015kwd,Buras:2015jaq,Tanimoto:2016yfy,Buras:2016dxz,Kitahara:2016otd,Endo:2016aws,Bobeth:2016llm,Cirigliano:2016yhc,Endo:2016tnu,Bobeth:2017xry,Crivellin:2017gks,Bobeth:2017ecx,Haba:2018byj,Buras:2018lgu,Chen:2018ytc,Chen:2018vog,Matsuzaki:2018jui,Haba:2018rzf,Aebischer:2018rrz,Aebischer:2018quc,Aebischer:2018csl}.

To explore  new physics contributions to the  $\epsilon'/\epsilon$ and the Kaon indirect CP violation $\epsilon_K$, in this work, we investigate the diquark effects, where  the diquark is a colored scalar and can originate  from grand unified theories (GUTs)~\cite{Barr:1986ky,Barr:1989fi}. Even without GUTs, basically, a diquark is allowed in the $SU(3)_C\times SU(2)_L\times U(1)_Y$ gauge symmetry, and its representation in the symmetry group depends on the coupled quark-representation~\cite{Assad:2017iib}. In this study, we  concentrate on the color triplet and $SU(2)_L$ singlet diquark.

 Although the diquark effects on $\epsilon_K$ and $\epsilon'/\epsilon$ were investigated in~\cite{Barr:1989fi}, some new diquark characteristics are found in this study, which can be summarized  as follows: (a) the $SU(2)_L$ singlet diquark can couple to the left-handed doublet  and  right-handed singlet quarks simultaneously. (b) When the sizable top-quark mass is taken into account, the $\Delta S=2$ box diagrams with the intermediates of $W$-boson (including charged Goldstone boson) and  diquark become significant, in which the effects were ignored in ~\cite{Barr:1989fi}. (c) New scalar-scalar and tensor-tensor operators for $\Delta S=1$ are induced at the tree level; due to  large mixings between the scalar and tensor operators, the $\epsilon'/\epsilon$ is dominated by the isospin $I=2$ amplitude, which is produced by the tensor-tensor operators~\cite{Aebischer:2018rrz}. (d)   QCD and EW penguin diagrams  are included in $\epsilon'/\epsilon$, and with the renormalization group (RG) effect, it is found that the $I=2$ amplitude, induced by the $Q_8$ operator, become dominant. (e) Chromomagnetic  operators (CMOs) generated from the gluon-penguin diagrams are considered based on the matrix elements obtained in~\cite{Buras:2018lgu}.  
 
 Although the involved new free parameters generally can carry CP phases, in this work, we assume that the origin of the CP violation is still from the Kobayashi-Maskawa (KM) phase of the CKM matrix. This assumption can be removed if necessary.  Hence,  it can be concluded that $\epsilon'/\epsilon$ can be significantly enhanced by the diquark effects when the bound from $\epsilon_K$ is satisfied.  In addition, since rare $B$-meson processes, such as $B^0_q -\bar B^0_q$ ($q=d, s$) mixings, involve different parameters,  e.g. $g^{R,L}_{33}$, which are irrelevant to the current study, we do not discuss the $B$-meson physics in this study.

The paper is organized as follows: In Section II, we introduce the diquark Yukawa couplings to the SM quarks and gauge couplings to the gluons, $\gamma$, and $Z$-boson. In Section III, we derive the diquark-induced effective Hamiltonian for the $\Delta S=1$ and $\Delta S=2$ processes, where the used three-point vertex functions of $d\to s g^{(*)}, \gamma^{(*)}, Z^{(*)}$ are derived in the appendix. The hadronic effects for the $K\to \pi \pi$ decays and the  $K^0-\bar K^0$ transition are shown in Section IV. We also summarize the formulations of $\epsilon'/\epsilon$ and $\epsilon_K$ from various operators in this section.  The constraints from  $\Delta S=2$  are shown in Section V.  The detailed numerical analysis on $\epsilon'/\epsilon$ based on various different  mechanisms is given in Section VI.  A summary is given in Section VII.

\section{Color-triplet diquark Yukawa and gauge couplings}

In this section, we introduce  the diquark Yukawa  couplings and  gauge couplings to the gauge bosons, including the gluons, photon, and $Z$-boson. Based on $SU(3)_C$ gauge invariance, it can be seen that the involving diquarks from the Yukawa sector can be color-triplet  and -sextet due to $3 \times 3 = \bar 3 + 6$. From the $SU(2)_L$ gauge invariance, the diquark candidates can be the $SU(2)_L$ singlet and triplet~\cite{Barr:1989fi}.   In order to provide a detailed study on the diquark effects,  we thus focus on the $SU(2)_L$ singlet and color-triplet diquark~\cite{Barr:1989fi}.

 It can be found that the possible diquark candidates in the $SU(3)_C\times SU(2)_L\times U(1)_Y$ gauge group are  $(\bar 3, 1, 1/3)$ and $(\bar 3, 1, -2/3)$.  For $(\bar 3,1,-2/3)$, the Yukawa couplings to the quarks are:
  \begin{equation}
  f_{ij} d^T_i C P_R {\bf H}^\dagger_3 d_j + H.c.\,,
  \end{equation}
where $C=i\gamma^2 \gamma^0$ is the charge conjugation; $P_{R(L)}=(1\pm \gamma_5)/2$, and  $f_{ij} = -f_{ji}$ due to $d^T_j\, C\, P_R {\bf H}^\dagger_3 d^T_i = - d^T_i\, C \, P_R {\bf H}^\dagger_3 d_j$. As a result,  the $\Delta S=2$ process and $\epsilon'/\epsilon$ both arise from one-loop effects. Thus, it may not be possible to explain the $\epsilon'/\epsilon$ data when the parameters are constrained by $\epsilon^{\rm exp}_K$. In addition, since the involved quarks inside the loop are the down-type quarks, due to no heavy quark enhancement, e.g. $m^2_t/m^2_{H_3}$, the effects are expected to be relatively small.  Hence, in this work, we devote ourselves to the  ${\bf H}_{3}(\bar 3,1,1/3)$ contributions to the $\Delta S=1$ and $\Delta S=2$ processes.  

\subsection{Yukawa couplings}

 The gauge invariant   Yukawa couplings of ${\bf H}_3(\bar 3,1,1/3)$ to the quarks in  the SM gauge symmetry can be written as:
 \begin{equation}
 -{\cal L}_{Y} = f_{ij} Q^T_{i} C \pmb{\varepsilon} {\bf H}^\dagger_{3} P_L Q_j + g^R_{ij} u^T_i C {\bf H}^\dagger_{3} P_R d_j + H.c, \label{eq:L_Y}
 \end{equation}
 where the indices $i, j$ denote the flavor indices; $\pmb{\varepsilon}$ is a $2\times 2$ antisymmetric matrix with  $\pmb{\varepsilon}_{12}=-\pmb{\varepsilon}_{21}=1$, and the representation of color-triplet diquark in $SU(3)_C$ can be expressed as ${\bf H}_{3} = K^a H^a_{3}$ with $(K^a)^{ij} = 1/\sqrt{2} \epsilon^{aij}$. For the complex conjugate state, we use $(\bar K_a)_{ij}=(K^a)^{ji}$, i.e.  ${\bf H}^\dagger_3= \bar K_a H^{*}_{3a}$; thus, we obtain $Tr (K^a \bar K_b)=\delta^a_b$ and $(K^a)^{\beta \alpha} (\bar K_a)_{\rho\sigma}=1/2(\delta_\sigma^{\beta}  \delta_\rho^{\alpha}-\delta_\rho^{\beta}\delta_\sigma^{\alpha})$. The explicit matrix forms of $K^a$ ($a=1,2,3$) can be found in~\cite{Han:2009ya}. From Eq.~(\ref{eq:L_Y}),  the color-gauge transformation of ${\bf H_3}$  in $SU(3)_C$  follows:
  \begin{equation}
  {\bf H}'_3 = U {\bf H}_3 U^T\,. \label{eq:H_3}
  \end{equation}
  If we decompose the $SU(2)_L$ quark doublet, the left-handed quark couplings can be formed as:
   \begin{equation}
   f_{ij} u^T_i C {\bf H}^\dagger_3 P_L d_j - f_{ji} d^T_j C {\bf H}^\dagger_3 P_L u_i = (f_{ij} + f_{ji} ) u^T_i C {\bf H}^\dagger_3 P_L d_j \,,
   \end{equation} 
 where the flavor indices $i, j$ do not sum. From the result, it can be seen that the color-triplet diquark Yukawa couplings to  the left-handed quarks are symmetric in flavor space, i.e. $g^L_{ij} \equiv f_{ij} + f_{ji} = g^L_{ji}$.  Using the new coupling definition, Eq.~(\ref{eq:L_Y}) can be written as:
  \begin{align}
 -{\cal L}_{Y} =u^T_{i} C \bar K_a \left( g^L_{ij} P_L + g^R_{ij} P_R \right) d_j  H^*_{3a} + H.c. \label{eq:Yukawa}
 \end{align}
We will use the Yukawa couplings $g^{L,R}_{ij}$ to show the diquark effects. 
  
\subsection{ Gluon couplings}

In order to calculate the gluon-penguin diagrams for the $d\to s g^{(*)}$ transition, we need to know the gluon couplings to the diquark.
%, where the interactions arise from the kinetic term of the diquark. 
Since the diquark state carries two color indices, the associated gauge covariant derivative will be different from that of  fundamental representation of $SU(3)_C$.  To find the covariant derivative of ${\bf H}_{3}$ in  $SU(3)_C$, we first consider the gauge transformation of $\partial_\mu {\bf H}_3$. 
Using  Eq.~(\ref{eq:H_3}) and $U=\exp(i g_s \pmb{\alpha}) = \exp(ig_s \alpha^a T^a)$, in which $T^a=\lambda^a/2$ and $\lambda^a$ are the Gell-Mann matrices, the color-gauge transformation of $\partial_\mu {\bf H}_3$ can be expressed as:
 \begin{equation}
 \partial_\mu {\bf H}'_{3} = U\left( \partial_\mu  {\bf H}_3 + ig_s \partial_\mu \pmb{\alpha} {\bf H}_3 + ig_s {\bf H}_3 \partial_\mu \pmb{\alpha}^T \right) U^T\,.
 \end{equation}
 It can be seen that there are two terms related to $\partial_\mu \pmb{\alpha}$; that is, $(\partial_\mu \pmb{\alpha}) {\bf H}_3$ and  $ {\bf H}_3 \partial_\mu \pmb{\alpha}^T$. 
From the result,  we can define the covariant derivative of ${\bf H}_3$, which transforms as ${\bf H}_{3}$ in $SU(3)_C$ symmetry, as:
 \begin{equation}
 D_\mu {\bf H}_3 \equiv  \partial_\mu {\bf H}_3 + i g_s {\bf A}_\mu {\bf H}_3 + ig_s {\bf H}_3 {\bf A}^T_\mu\,,
 \end{equation} 
 where ${\bf A}_\mu = T^a A^a_\mu$ denotes the gluon fields, and its gauge transformation is given by ${\bf A}'_\mu = U {\bf A}_\mu U^\dagger - \frac{i}{g_s} U \partial_\mu U^\dagger$.  We have checked that the identity $D'_\mu{\bf H}'_3 = U D_\mu {\bf H}_{3} U^T$ is satisfied under the $SU(3)_C$ transformation.
 
After finding $D_\mu {\bf H}_3$, the $SU(3)_C$ gauge invariant kinetic term of ${\bf H}_3$ can thus  be written and expanded as:
  \begin{align}
  Tr (D_\mu {\bf H}_3)^\dagger (D^\mu {\bf H}_3) & = \partial_\mu H^{*}_{3a} \partial^\mu H^a_3 + i g_s Tr \partial_\mu {\bf H}^\dagger_3 \left( {\bf A}_\mu {\bf H}_3 + {\bf H}_3 {\bf A}^T_\mu \right) \nonumber \\
  & -i g_s  Tr  \left( {\bf H}^\dagger _3 {\bf A}_\mu  +  {\bf A}^T_\mu {\bf H}^\dagger_3  \right)\partial_\mu {\bf H}_3 \nonumber \\
  & + g^2_s Tr  \left( {\bf H}^\dagger _3 {\bf A}_\mu  +  {\bf A}^T_\mu {\bf H}^\dagger_3  \right) \left( {\bf A}_\mu {\bf H}_3 + {\bf H}_3 {\bf A}^T_\mu \right)\,.
  \end{align}
We can read out the gluon couplings to the diquark-pair from the second and third terms, where their color factors can be factored out as:
 \begin{align}
 {\cal L}_{{\bf A}H_3 H_3} & = i g_s \left( Tr \bar K_a T^A K^b + Tr \bar K_a K^b (T^A)^T \right) (\partial^\mu  H^*_{3a})  H^b_3 A^A_\mu  \nonumber \\
  & -   i g_s \left( Tr \bar K_a T^A K^b + Tr \bar K_a K^b (T^A)^T \right) H^*_{3a}  (\partial^\mu H^b_3) A^A_\mu\,.
 \end{align}
It can be easily shown that $Tr \bar K_a  K^b (T^A)^T = Tr \bar K_a T^A K^b$, and the interaction of ${\bf A}_\mu H_3 H_3$ can then be rewritten as:
\begin{align}
{\cal L}_{{\bf A} H_3 H_3} =i  g_s  (t^A)^b_a \left[ (\partial^\mu H^*_{3a}) H^b_3 A^{A}_{\mu}  -  H^*_{3a} (\partial^\mu H^b_3 ) A^{A}_{\mu} \right]\,,
\end{align}
with $(t^A)^{b}_{a} = 2 Tr (\bar K_a T^A K^b)$. As a result, the associated Feynman rule can be obtained as: 
 \begin{equation}
 A^A_\mu  H^*_{3a} H^b_3 : -i g_s (t^A)^b_a ( p_b + p_a)_\mu \,. \label{eq:gH3H3}
 \end{equation}
Additionally,  the  color trace factors $Tr \bar K_a T^A K^b$ indeed are related  to the generators of $SU(3)_C$, and the relationship can be built as follows:
 \begin{align}
 Tr \bar K_a T^A K^b  & = (\bar K_a)_{\alpha \beta} (T^A)_{\rho}^\beta (K^b)^{\rho\alpha} = \frac{1}{2} \epsilon_{a \beta \alpha} \epsilon^{b \rho\alpha}  (T^A)_{\rho}^\beta  \nonumber \\
 &= \frac{1}{2} \left( \delta_{a}^{b} Tr (T^A )- (T^A)^b_a \right) = - \frac{(T^A)^b_a}{2}\,,
 \end{align}
 where  $\epsilon_{ijk} \epsilon^{\ell m k} = \delta_{i}^{\ell} \delta_{j}^{m} - \delta_{i}^{m}\delta_{j}^{\ell}$ and $Tr(T^A)=0$ are used.

\subsection{Photon and $Z$-boson gauge couplings}

Since ${\bf H}_3$ is an $SU(2)_L$ singlet, the ${\bf H}_3$ hypercharge is equal to its electric charge. In order to know the photon and $Z$-boson gauge couplings to the diquark, we write the $U(1)_Y$ covariant derivative of ${\bf H}_3$ as:
 \begin{equation}
 D_\mu {\bf H}_3 = (\partial_\mu + i g' Y_{H_3} B_\mu ){\bf H}_3\,,
 \end{equation}
 where $g'$ is the $U(1)_Y$ gauge coupling constant; $Y_{H_3}$ is the ${\bf H}_3$ hypercharge, and $B_\mu$ is the $U(1)_Y$ gauge field. 
The $U(1)_Y$ gauge invariant kinetic term of ${\bf H}_3$ can  then be expressed as:
 \begin{align}
 Tr (D_\mu {\bf H}_3)^\dagger(D^\mu {\bf H_3}) & = (D_\mu H^*_{3a})(D^\mu H^a_{3}) = \partial_\mu H^*_{3a} \partial^\mu H^a_3  \nonumber \\
 &+ i g' Y_{H_3} \left( \partial_\mu H^*_{3a} H^a_3 - H^*_{3a} \partial_\mu H^a_3 \right)  B^\mu \nonumber \\
 & + g'^2 Y^2_{H_3} B^2 H^*_{3a} H^a_{3}\,,
 \end{align}
 where $Tr \bar K_a K^b=\delta^b_a$ has been applied to the first equality. Using $B_\mu = \cos\theta_W A_\mu - \sin\theta_W Z_\mu$, the EW  gauge couplings to the diquark can be obtained as:
  \begin{align}
  {\cal L}_{VH_3 H_3}  & = i e_{H_3} e (\partial_\mu H^*_{3a} H^a_{3} - H^*_{3a} \partial_\mu H^a_3 )A^\mu \nonumber \\
  & -i  \frac{g e_{H_3} \sin^2\theta_W}{\cos\theta_W} (\partial_\mu H^*_{3a} H^a_{3} - H^*_{3a} \partial_\mu H^a_3 )Z^\mu\,, \label{eq:H3-AZ}
  \end{align} 
  where $\theta_W$ is the Weinberg's angle; $e=g'\cos\theta_W = g\sin\theta_W$ and $g'/g=\tan\theta_W$ are applied; $g$ is the $SU(2)_L$ gauge coupling constant, and $e_{H_3}=Y_{H_3}=1/3$ is the $H^a_{3}$ electric charge. The associated Feynman rule can be obtained as: 
 \begin{align}
 A_\mu  H^*_{3a} H^b_3 & : -i e_{H_3} e ( p_b + p_a)_\mu  \delta^b_a \,, \label{eq:photonH3H3}   \\
 Z_\mu H^*_{3a} H^b_3  & : i \frac{g e_{H_3} \sin^2\theta_W}{\cos\theta_W} ( p_b + p_a)_\mu  \delta^b_a \,. \label{eq:ZH3H3} 
 \end{align}

\section{Diquark-induced effective Hamiltonian for the $\Delta S=1$ and  $\Delta S=2$ processes}

In the diquark model, the $K\to \pi \pi$ decays can be produced through the tree, QCD penguin, and EW penguin diagrams. In this section, we discuss  in detail the effective Hamiltonian for the $\Delta S=1$ processes induced by each type of Feynman diagrams. For the $\Delta S=2$ process, the involved effects include one $W$ and one ${\bf H}_3$ box diagram and pure ${\bf H}_3$-mediated box diagram. Since the  Yukawa couplings of the ${\bf H}_{3}$ to the light quarks are strictly constrained by the tree processes,  we assume that the Yukawa couplings  related to  the third generation quarks are not suppressed and can be relatively large, e.g., $g^{L(R)}_{31,32}\lesssim 0.1$. Therefore,  we only consider the top-quark box diagrams and directly neglect the light-quark boxes. 

\subsection{ Effective Hamiltonian for $K \to \pi \pi$}

\subsubsection{\bf Tree diagram}

\begin{figure}[phtb]
\includegraphics[scale=0.8]{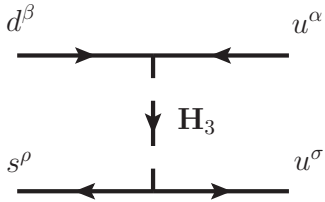}
 \caption{ Tree diagram  for the $K \to \pi \pi$ decays mediated by color-triplet diquark ${\bf H}_{3}$. }
\label{fig:tree}
\end{figure}

 The Feynman diagram of tree-level diquark contribution to the $K\to \pi \pi$ decays is shown in Fig.~\ref{fig:tree}. 
   %%%%
 Using the Yukawa couplings in Eq.~(\ref{eq:Yukawa}), the four-fermion interactions can be written as:
 \begin{align}
{\cal H}_{\rm tree} & = - \frac{1 }{2m^2_{H_3}} \left[ g^L_{11} g^{L*}_{12} (\overline{u^{C\alpha}} P_L d^\beta)\, (\bar{s}_\beta  P_R u^{C}_{\alpha}) 
+ g^R_{11} g^{R*}_{12} (\overline{u^{C\alpha}} P_R d^\beta)\, (\bar s_\beta  P_L u^{C}_{\alpha}) \right.\nonumber  \\
& \left. + g^L_{11} g^{R*}_{12} (\overline{u^{C\alpha}} P_L d^\beta)\, (\bar s_\beta  P_L u^{C}_{\alpha} ) 
+ g^R_{11} g^{L*}_{12} (\overline{u^{C \alpha}} P_R d^\beta)\, (\bar s_\beta  P_R u^{C}_{\alpha}) 
\right]\,, \label{eq:H_tree}
 \end{align}
 where the charge-conjugation state of a fermion is defined by $f^C = C \gamma_0 f^* = C \bar{f}^T$. We can express the ${\cal H}_{\rm tree}$ in terms of fermion states using the Fierz and $C$-parity transformations, which are:
  \begin{align}
  \bar f_3 P_\chi f_2 \, \bar f_1 P_\chi f_4 & = - \frac{1}{2} (\bar f_2 P_\chi f_1 ) (\bar f_{3} P_\chi f_4) - \frac{1}{8} (\bar f_2 \sigma_{\mu \nu} P_\chi f_1 ) (\bar f_{3} \sigma^{\mu\nu} P_\chi f_4)\,, \nonumber \\
  %
%  \bar f_3 \sigma_{\mu \nu} P_\chi f_2 \,\bar f_{1} \sigma^{\mu\nu} P_\chi f_4 & = -6 \bar f_2 P_\chi f_1 \bar f_3 P_\chi f_4 + \frac{1}{2} \bar f_2 
%\sigma_{\mu\nu} P_\chi f_1 \bar f_3 \sigma^{\mu \nu} P_\chi f_4\,, \nonumber \\
  %
  \overline{f^C} P_\chi f^C &= \bar f  P_\chi f \,, \nonumber \\
   \overline{f^C} \sigma_{\mu \nu} P_\chi f^C &= - \bar f  \sigma_{\mu\nu} P_\chi f\,, \label{eq:FT}
  \end{align}
  with $P_\chi=P_{R(L)}$. As a result, Eq.~(\ref{eq:H_tree}) can be formulated as:
  \begin{align}
  {\cal H}_{\rm tree} & = -\frac{G_F V^*_{ts} V_{td}}{ \sqrt{2}}  \frac{y_W}{2} \left[ \zeta^{LL}_{21}\left( Q_1 - Q_2 \right) + \zeta^{RR}_{21} \left( Q'_{1} - Q'_{2} \right) \right. \nonumber \\
  & -  \zeta^{LR}_{21} \left( 4 \left(Q^{SLL,u}_1 + Q^{SLL,u}_2 \right)+ Q^{SLL,u}_3 + Q^{SLL,u}_{4} \right) \nonumber \\
 &\left.  - \zeta^{RL}_{21} \left( 4\left( Q'^{SLL,u}_1 + Q'^{SLL,u}_2 \right)+ Q'^{SLL,u}_3 + Q'^{SLL,u}_{4} \right) \right] \,, \label{eq:treeH}
  \end{align}
 where  $G_F$ is the Fermi constant; $V_{ij}$ denotes the CKM matrix element; $y_W=m^2_W/m^2_{H_3}$, and  the parameters $\zeta^\chi_{21}$ are defined as:
  \begin{equation}
  \zeta^{LL(RR)}_{21} = \frac{g^{L(R)}_{11} g^{L(R)*}_{12} }{g^2 V^*_{ts} V_{td}}\,, ~ \zeta^{LR(RL)}_{21} = \frac{g^{L(R)}_{11} g^{R(L)*}_{12}}{g^2 V^*_{ts} V_{td}}\,.
  \end{equation}
    Following the notations shown in~\cite{Buras:2000if,Aebischer:2018rrz},  the effective operators are defined as:
  \begin{align}
  Q_1 & =( \bar s d)_{V-A} (\bar u u)_{V-A}\,,~ Q_2  = ( \bar s u)_{V-A} (\bar u d)_{V-A} \,,\nonumber \\
  %
 % Q'^u_3 & =( \bar s d)_{V+A} (\bar u u)_{V+A}\,,~ Q'^u_4  = ( \bar s_\alpha  d^\beta)_{V+A} (\bar u_\beta u^\alpha)_{V+A} \nonumber \\
%
Q^{SLL,u}_{1} & = ( \bar s_\alpha P_L u^\beta) (\bar u_\beta P_L d^\alpha)\,, ~Q^{SLL,u}_{2} = ( \bar s_\alpha P_L d^\alpha) (\bar u_\beta P_L u^\beta)\,, \nonumber \\
Q^{SLL,u}_{3} & =  -( \bar s_\alpha \sigma_{\mu\nu} P_L u^\beta) (\bar u_\beta \sigma^{\mu\nu} P_L d^\alpha)\,,~
Q^{SLL,u}_4 = -( \bar s_\alpha \sigma_{\mu\nu}P_L d^\alpha ) (\bar u_\beta \sigma^{\mu\nu}P_L u^\beta )\,,
  \end{align}
where $(\bar f f)_{V-A}=\bar f \gamma_{\mu}(1-\gamma_5) f$, and the prime operators  can be obtained from unprimed ones using  $P_{L(R)}$ instead of  $P_{R(L)}$. It can be seen that the current-current interactions induced at the tree-level involve vector-, scalar-, and tensor-type currents.  
Although  the tensor-tensor operator contributions to the $K\to \pi \pi$ decays  vanish at the factorization scale,  since a large mixing between the scalar-scalar and tensor-tensor operators is induced at  one-loop QCD corrections~\cite{Aebischer:2018rrz}, the tensor-type interaction can have a large contribution to $\epsilon'/\epsilon$.

\subsubsection{\bf QCD penguins}

\begin{figure}[phtb]
\includegraphics[scale=0.5]{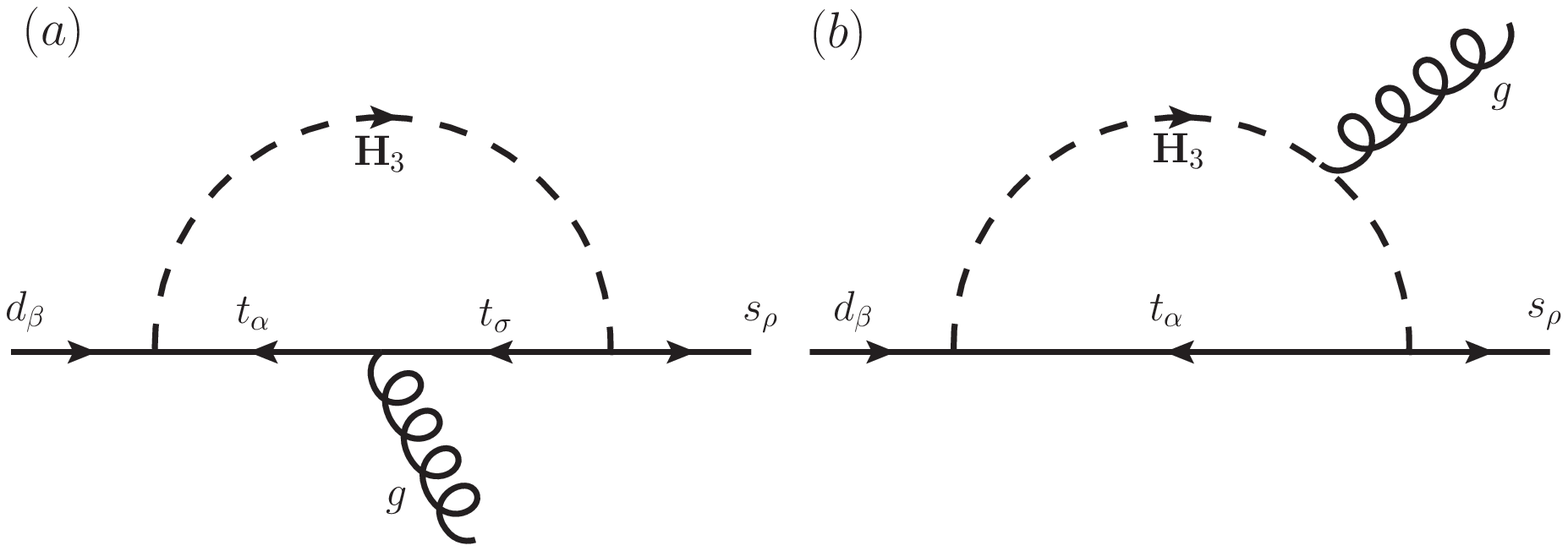}
 \caption{ Gluon-penguin diagrams for the  $d\to s g^{(*)}$ transition mediated by color-triplet diquark ${\bf H}_{3}$. }
\label{fig:penguin}
\end{figure}

In addition to the tree-level diagrams, the $K\to \pi \pi$ decays in the diquark model can arise from the gluon-penguin diagrams, as shown in Fig.~\ref{fig:penguin}.  As is known, the loop diagram usually leads to an ultraviolet divergence. To obtain the finite coupling for the $d \to s g^{(*)}$ vertex, we have to renormalize the three-point vertex function by including the self-energy diagram for the $d\to s$ flavor changing transition. The detailed discussions for renormalizing  the $d \to s g^{(*)}$ vertex are given in the appendix; here, we simply use the obtained results of Fig.~\ref{fig:penguin}(a) and (b) to produce the effective Hamiltonian for the $K\to \pi \pi$ decays. 

   %%%%

Because the gluon momentum $k$ satisfies  $k^2\ll m^2_t, m^2_{H_3}$, we can  expand the three-point functions in terms of $k^2 /m^2_{H_3}$ and keep the leading $k^2/m^2_{H_3}$ terms. Thus, based on the renormalized  vertex obtained in Eq.~(\ref{eq:RVg}),  the penguin-induced Lagrangian for $d \to s g^*$  can be expressed as:
 \begin{equation}
 {\cal L}_{d \to s g^*} = - \frac{g_s\, k^2 }{ (4\pi)^2 m^2_{H_3}} I_{G1}(y_t) \bar s \gamma^\mu \left( g^L_{31}  g^{L*}_{32}P_L + g^R_{31}  g^{R*}_{32} P_R \right) T^a d \, A^{a}_{\mu}\,,
 \end{equation}
 where $I_{G1}(y_t)$ with $y_t=m^2_t/m^2_{H_3}$ denotes the loop integral function and can be found from Eq.~(\ref{eq:loop_fg}). The $k^2$ factor in the numerator can be used to cancel the off-shell gluon propagator, i.e.,  $1/(k^2 + i \varepsilon)$.
Accordingly, the effective Hamiltonian for the $d \to s \bar q q$ decays from the gluon-penguin can be obtained as:
\begin{align}
{\cal  H}_{\rm QCD} &= - \frac{\alpha_s I_{G1}(y_t)} {32 \pi m^2_{H_3}} \left[ g^{L*}_{32} g^L_{31} \left( Q_4 + Q_6 - \frac{1}{3} 
 Q_3 - \frac{1}{3} Q_5 \right) \right. \nonumber \\
&\left. +  g^{R*}_{32} g^R_{31} \left( Q'_4 + Q'_6\ -  \frac{1}{3} 
Q'_3 - \frac{1}{3} Q'_5 \right) \right]\,,
\end{align}
 where we have used:
  \begin{equation}
  (T^a)^\alpha_\beta (T^a)^\rho_\sigma = \frac{1}{2} \left( \delta^\alpha_\sigma \delta^\rho_\beta - \frac{1}{3} \delta^\alpha_\beta \delta^\rho_\sigma \right)\,;
  \end{equation}
 the unprimed operators at the $m_{H_3}$ scale are the same as those in the SM and can be found as:
 \begin{align}
 Q_3 & = (\bar s d)_{V-A} \sum_q (\bar q q)_{V-A}\,,~ Q_4 =  (\bar s_\alpha d^\beta)_{V-A} \sum_q (\bar q_\alpha q^\beta)_{V-A}\,, \nonumber \\
  Q_5 & = (\bar s d)_{V-A} \sum_q (\bar q q)_{V+A}\,,~ Q_6 =  (\bar s_\alpha d^\beta)_{V-A} \sum_q (\bar q_\alpha q^\beta)_{V+A}\,, \label{eq:QCD-penguin}
 \end{align}
and the prime operators can be obtained from the unprimed ones via the exchange of $P_{L(R)}$ and  $P_{R(L)}$.

\subsubsection{\bf EW penguins}

\begin{figure}[phtb]
\includegraphics[scale=0.5]{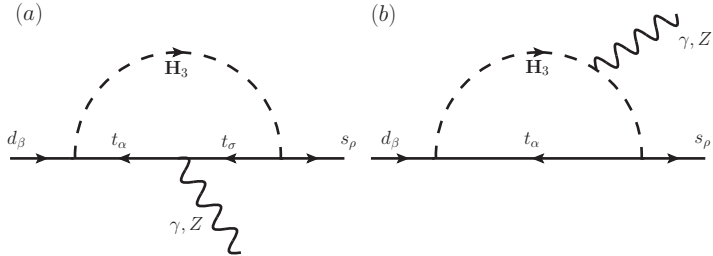}
 \caption{ Feynman diagrams for the $d \to s (\gamma^{(*)}\,, Z^*)$ processes. }
\label{fig:AZ-penguin}
\end{figure}

The $d \to s q \bar q$ decays can be also induced from the EW penguin diagrams through the mediation of the off-shell photon and $Z$-boson, where the Feynman diagrams are shown in Fig.~\ref{fig:AZ-penguin}.  Similar to the case in $d\to s g^{(*)}$, there are ultraviolet divergences in the loop integrals of Fig.~\ref{fig:AZ-penguin}(a) and (b). The discussions for the  divergence cancellation are given in the appendix. According to Eqs.~(\ref{eq:L_gamma}) and (\ref{eq:L_Z}),  the  loop-induced Lagrangian for $d\to s (\gamma^*, Z^*)$ can be written as:
  \begin{align}
  {\cal L}_{s\to d \gamma^*, Z^*} & = - \frac{e k^2}{3 (4\pi)^2 m^2_{H_3}} I_{\gamma 1} (y_t) \bar s \gamma^\mu  \left(g^L_{31} g^{L*}_{32}  P_L + g^R_{31} g^{R*}_{32}  P_R \right)  d \, A_\mu\,, \nonumber \\
&  - \frac{g }{2 \cos\theta_W (4\pi)^2 } \bar s \gamma^\mu I_{Z}(y_t) \left( g^L_{31} g^{L*}_{32}  P_L  - g^R_{31} g^{R*}_{32}  P_R \right) d\, Z_\mu\,, \label{eq:LdsgaZ}
  \end{align}
  where $I_{\gamma 1}$ and $I_{Z}$ are  the associated loop functions and can be found in Eqs.~(\ref{eq:I_gamma}) and (\ref{eq:I_Z}).
  
  %%%%

 Based on Eq.~(\ref{eq:LdsgaZ}),  the effective Hamiltonian for the $d\to s q \bar q$ decays can be written as:
\begin{equation}
{\cal H}_{EW}  = - \frac{G_F V^*_{ts} V_{td} }{\sqrt{2}} \left[ C^{Z}_3 Q_3 + C'^Z_5 Q'_5 + \sum^{10}_{i=7} \left( C^{\gamma Z}_i Q_i + C'^{\gamma Z}_i  Q'_i\right) \right]\,, 
% + C^{\gamma Z}_8 Q_8 + C^{\gamma Z}_9 Q_9  \right. \nonumber \\
%& \left.  +C^{\gamma Z}_{10} Q_{10}+ C'^{\gamma Z}_7 Q'_7 + C'^{\gamma Z}_8 Q'_8  + C'^{\gamma Z}_9 Q'_9 +C'^{\gamma Z}_{10} Q'_{10}
 %\right)\,,
\end{equation}
where  the effective operators $Q_7$-$Q_{10}$ are the same as those in the SM and are expressed as:
  \begin{align}
  Q_7 & =  \frac{3}{2} (\bar s d)_{V-A} \sum _q  e_q (\bar q q)_{V+A}\,, \ Q_8 =  \frac{3}{2} (\bar s_\alpha d^\beta)_{V-A} \sum _q e_q (\bar q_\beta q^\alpha)_{V+A}\,, \nonumber \\
   Q_9 & =  \frac{3}{2} (\bar s d)_{V-A} \sum _q  e_q (\bar q q)_{V-A}\,, \ Q_{10} =  \frac{3}{2} (\bar s_\alpha d^\beta)_{V-A} \sum _q e_q (\bar q_\beta q^\alpha)_{V-A}\,,
  \end{align}
 and $e_q$ is the $q$-quark electric charge. The prime operators $Q'_7$-$Q'_{10}$ can be obtained from the unprimed operators through the exchange of $P_{L(R)}$ and  $P_{R(L)}$. The effective Wilson coefficients $C^{(\prime)Z}_i$ and $C^{(\prime)\gamma Z}_i$ are  given as:
 \begin{align}
 C^Z_3 & = \frac{\alpha}{6\pi \sin^2 \theta_W}  \frac{I_{Z}(y_t) h^L_{21} }{4 }  \,, ~ C'^Z_5 = - \frac{\alpha}{6\pi \sin^2 \theta_W}  \frac{I_{Z}(y_t) h^R_{21} }{4} \,, \nonumber \\
  C^{\gamma Z}_7 & = \frac{4 \alpha}{6\pi }  \frac{I_{Z}(y_t) h^L_{21} }{4 } +  \frac{\alpha}{6\pi}  \frac{ 2 y_W I_{\gamma 1}(y_t) h^L_{21} }{3 } \,, ~ C^{\gamma Z}_9 = C^{\gamma Z}_7 - 4 C^Z_3  \,, \nonumber \\
  C'^{\gamma Z}_9 &=  -\frac{4 \alpha}{6\pi} \frac{I_{Z}(y_t) h^R_{21}}{4}+ \frac{\alpha}{6\pi} \frac{2 y_W I_{\gamma 1}(y_t)  h^R_{21} }{3 }\,, ~ C'^{\gamma Z}_7 = C'^{\gamma Z}_9 - 4 C'^Z_5 \,,\label{eq:WC_Q3-6}
 \end{align}
where $\alpha=e^2/4\pi$; $y_W = m^2_W/m^2_{H_3}$; $C^{(\prime)\gamma Z}_8=C^{(\prime) \gamma Z}_{10}=0$, and the $h^{L,R}_{21}$ parameters are defined by:
 \begin{equation}
 h^{L}_{21} = \frac{g^{L*}_{32} g^{L}_{31} }{g^2 V^*_{ts} V_{td} }\,, ~ h^{R}_{21} = \frac{g^{R*}_{32} g^{R}_{31} }{g^2 V^*_{ts} V_{td} }\,. \label{eq:hLR_21}
 \end{equation}
 We can use the new  parameters $h^{L,R}_{21}$ to study the diquark contributions to $\epsilon'/\epsilon$. 

\subsubsection{\bf Combination of the QCD and EW penguins and CMOs}

After respectively obtaining the QCD and EW penguin contributions to the $d \to s q \bar q$ decays, the effective Hamiltonian for the $\Delta S=1$ processes in the diquark model can be combined as:
 \begin{equation}
 {\cal H}_{\Delta S=1}  = - \frac{ G_F V^*_{ts} V_{td} }{\sqrt{2} }   \sum_{i=3}^{10} \left( y^{H_3} _i Q_i  + y'^{H_3}_i Q'_i \right)\,, \label{eq:QCD_EW}
 \end{equation}
 where the effective Wilson coefficients $y^{H_3}_i$ and $y'^{H_3}_i$ are given as:
 \begin{align}
 y^{H_3} _3 & =  - \frac{\alpha_s}{12\pi} h^L_{21} y_W I_{G1}(y_t) + C^Z_3\,, ~ y^{H_3}_4  = \frac{\alpha_s}{4\pi} h^L_{21} y_W I_{G1}(y_t) \,, \nonumber \\
 y^{H_3}_5 & = -\frac{\alpha_s}{12\pi} h^L_{21} y_W I_{G1}(y_t) \,, ~y^{H_3}_6  = y^{H_3}_4 \,, ~ y^{H_3}_7 = C^{\gamma Z}_7 \,, ~y^{H_3}_9  = C^{\gamma Z}_9 \,,
 \nonumber \\
 y'^{H_3}_3 & =  - \frac{\alpha_s}{12\pi} h^R_{21} y_W I_{G1}(y_t) \,,  ~ y'^{H_3}_4 = \frac{\alpha_s}{4\pi} h^R_{21} y_W I_{G1}(y_t)\,, \nonumber \\
 y'^{H_3}_5 & = y'^{H_3}_3 + C'^{Z}_5\,, ~ y'^{H_3}_6 = y'^{H_3}_4\,, ~ y'^{H_3}_7=C'^{\gamma Z}_7\,, ~ y'^{H_3}_9 = C'^{\gamma Z}_9\,, \label{eq:WCH3}
  \end{align}
 and $y^{H_3}_{8, 10}= y'^{H_3}_{8,10}=0$. Hence, we will use Eqs.~(\ref{eq:QCD_EW}) and (\ref{eq:WCH3}) to study $\epsilon'/\epsilon$. 
 
 In addition to the QCD and EW penguins, the gluonic and electromagnetic dipole operators can contribute to the $K\to \pi \pi$ decays. 
 Since the strong interactions dominate, we only study the gluonic dipole contributions in this paper. 
 %The electromagnetic contribution can be found from Eq.~(\ref{eq:L_gamma}) in the appendix. 
 Therefore, according to Eq.~(\ref{eq:RVg}), the  effective Hamiltonian for $d\to s g$ in the chromomagnetic dipole form can be written as: 
 \begin{equation}
 {\cal H}_{d\to s g}  = - \frac{G_F V^*_{ts} V_{td}}{\sqrt{2}} \left( C^{H_3}_{8G} Q_{8G} + C'^{H_3}_{8G} Q'_{8G} \right)\,,  \label{eq:CMD}
 \end{equation}
 where the dimension-6 CMOs $Q^{(\prime)}_{8G}$ are defined as:
 \begin{align}
 Q_{8G} &= \frac{ g_s}{8 \pi^2} m_s \bar s \sigma\cdot G P_L d\,, \nonumber \\
 Q'_{8G} &= \frac{g_s }{8 \pi^2 } m_d \bar s \sigma\cdot G P_R d\,,
 \end{align}
with $\sigma\cdot G = \sigma^{\mu \nu} T^a G^a_{\mu\nu}$, and the associated Wilson coefficients are shown as:
\begin{equation}
C^{H_3}_{8G} = \frac{m_t}{m_s} \frac{g^{R*}_{32}}{g^{L*}_{32}}h^L_{21} y_W I_{G2}(y_t) \,, ~ C'^{H_3}_{8G} = \frac{m_t}{m_d} \frac{g^{L*}_{32}}{g^{R*}_{32}}h^R_{21} y_W I_{G2}(y_t)\,. \label{eq:CH3_8G}
\end{equation}
 $I_{G2}$ is the loop integral function and can be found from Eq.~(\ref{eq:loop_fg}). Because the involved ${\bf H}_3$ Yukawa couplings in the induced CMOs are $g^{R*}_{32} g^{L}_{31}$ and $g^{L*}_{32} g^R_{31}$, from Eq.~(\ref{eq:CH3_8G}), it is seen that $h^{L}_{21}$ and $h^{R}_{21}$ are associated with $g^{R*}_{32}/g^{L*}_{32}$ and $g^{L*}_{32}/g^{R*}_{32}$ factors, respectively.  Since $g^R_{32}$ and $g^L_{32}$ cannot be singly constrained, we can take $g^R_{32}/g^L_{32}\approx 1$ and just use $h^{L,R}_{21}$ as the independent variables to study the  CMO effects. 
 Recently, the $K\to \pi \pi$ matrix elements of the   CMOs were calculated based on a DQCD approach~\cite{Buras:2018evv}, and the results are consistent with the  lattice QCD, as calculated by ETM collaboration~\cite{Constantinou:2017sgv}.  We  use the Hamiltonian in Eq.~(\ref{eq:CMD}) and the $K\to \pi \pi$ matrix elements obtained using the DQCD approach to investigate the CMO effects on $\epsilon'/\epsilon$.

\subsection{$\Delta S=2$ in the diquark model}

\begin{figure}[phtb]
\includegraphics[scale=0.75]{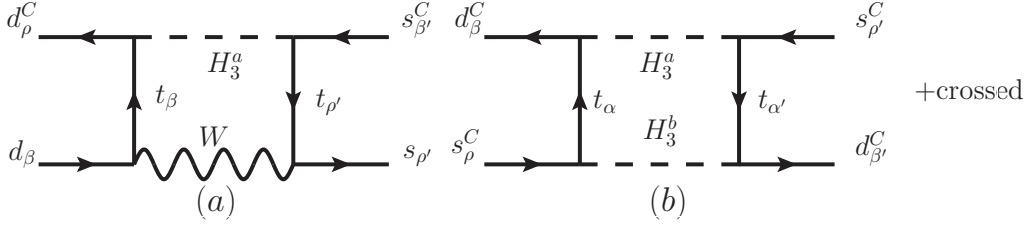}
 \caption{ Box diagrams for $\Delta S=2$ in the diquark model, where the subscripts denote the color indices.}
\label{fig:box_W_H3}
\end{figure}

Next, we study the ${\bf H}_3$ contributions to the $\Delta S=2$ process, where the involved Feynman diagrams are sketched in Fig.~\ref{fig:box_W_H3}. It has been  pointed out that the contribution of Fig.~\ref{fig:box_W_H3}(a) vanishes in the chiral limit, i.e., $m_t\sim 0$~\cite{Barr:1989fi}.  In the following analysis, in addition to discussing the origin of the vanished result, we also demonstrate that the  Fig.~\ref{fig:box_W_H3}(a) contribution is  interesting and important when $m_t\approx 165$ GeV and $m_{H_3} \approx {\cal O}$(1) TeV are taken. 

  %%%%

 To study the diquark contributions to $\Delta S=2$, we follow the notations in~\cite{Buras:2001ra} and write  the effective Hamiltonian  as:
 \begin{equation}
 {\cal H}_{\Delta S=2} = \frac{ G^2_F V_{\rm CKM} }{16 \pi^2} m^2_W \sum_{ i}    C^{\chi} _{i} (\mu) Q^{\chi}_{i} \,, \label{eq:H_DS2}
 \end{equation}
where $V_{\rm CKM}= (V^*_{ts} V_{td})^2 $ is the product of the CKM matrix elements; $C^\chi_i(\mu)$ are the Wilson coefficients at the $\mu$ scale,  and the relevant operators $Q^\chi_i$ are given as:
 \begin{align}
 Q^{VLL}_1 & = (\bar s \gamma_\mu P_L d) (\bar s \gamma^\mu P_L d)\,, \nonumber \\
 Q^{LR}_1 & = (\bar s \gamma_\mu P_L d) (\bar s \gamma^\mu P_R d)\,, \nonumber \\
 Q^{LR}_2 & = (\bar s P_L d) (\bar s P_R d)\,, \nonumber \\
 Q^{SLL}_1 & = (\bar s P_L d) (\bar s P_L d)\,, \nonumber \\
 Q^{SLL}_{2} & = (\bar s \sigma_{\mu\nu} P_L d) (\bar s \sigma^{\mu \nu} P_L d). \label{eq:Qs}
 \end{align}
The operators $Q^{VRR}_1$ and $Q^{SRR}_{i}$ can be obtained from $Q^{VLL}_1$ and $Q^{SLL}_{i}$ by switching $P_R$ and $P_L$, respectively. We use the effective operators in Eq.~(\ref{eq:Qs}) to show the diquark contributions. 

 \subsubsection{\bf Box diagrams with one $W$-boson and one diquark }
 
  Based on  the Yukawa couplings in Eq.~(\ref{eq:Yukawa}) and using the 't Hooft-Feynman gauge, the effective Hamiltonian  for $\Delta S=2$ via the mediation of $W$ and ${\bf H}_3$ shown in Fig.~\ref{fig:box_W_H3}(a) can be written as:
  \begin{align}
 {\cal H}^{WH_3}_{\Delta S=2} & = -  \frac{g^2 V^*_{ts} V_{td} }{2} (\bar K_a)_{\rho\beta} (K^a)^{\rho'\beta'} \int \frac{d^4 q}{(2\pi)^4 )} \frac{1}{(q^2-m^2_{H^2_3}) (q^2-m^2_W ) (q^2 - m^2_t)^2 } \nonumber \\
 & \times  \left[ g^L_{31} g^{L*}_{32}  \left(\overline{d^{C \rho}} \slashed{q} \gamma_\mu P_L d^\beta \right) \left( \bar s_{\rho'} \gamma^\mu \slashed{q} P_R s^C_{\beta'} \right) + m^2_t g^R_{31} g^{R*}_{32} \left( \overline{d^{C \rho}} \gamma_\mu P_L d^{\beta} \right) \left( \bar s_{\rho'} \gamma^\mu P_L s^C_{\beta'} \right) \right]\,,
  \label{eq:W_H3}
 \end{align}
It can be seen that because $W$-boson only couples to the left-handed quarks, without the chirality flipping effect, e.g. $m_t$, the first term  depends on $g^L_{31} g^{L*}_{32}$. With the chirality flip, which arises from  the mass insertions in the two top-quark propagators, the second term in Eq.~(\ref{eq:W_H3}) is associated with the right-handed quark couplings $m^2_t g^R_{31} g^{R*}_{32}$. 

Although $g^L_{31} g^{L*}_{32}$ appears in Eq.~(\ref{eq:W_H3}),   we demonstrate that its contribution  indeed vanishes when the color factor and Fierz transformation are considered together. 
 Using $\gamma_\mu \gamma_\nu = g_{\mu \nu} - i \sigma_{\mu \nu}$ and Fierz transformation, the first term in Eq.~(\ref{eq:W_H3}) can be derived as:
   \begin{align}
    \left(\overline{d^{C\rho}} \gamma_\nu \gamma_\mu P_L d^\beta \right) \left( \bar s_{\rho'} \gamma^\mu \gamma^\nu P_R s^C_{\beta'} \right) 
   & = -2 \left( \overline{d^{C\rho}} \gamma_\mu P_R s^C_{\beta'} \right) \left( \bar s_{\rho'} \gamma^\mu P_L d^\beta \right) \nonumber \\
   & = 2 \bar s_{\beta'} \gamma_\mu P_L d^\rho \bar s_{\rho'} \gamma^\mu P_L d^\beta\,. 
   \end{align}
   We note that because the chirality of initial quark can not match with that of final quark, the tensor-type current is not allowed.
   Combined with the color factor $(\bar K_a)_{\rho\beta} (K^a)^{\rho'\beta'} = (\delta^{\beta'}_{\rho} \delta^{\rho'}_{\beta} - \delta^{\rho'}_\rho \delta^{\beta'}_\beta)/2$, the result of above equation can be proceeded  as:
    \begin{equation}
    2 (\bar K_a)_{\rho\beta} (K^a)^{\rho'\beta'}  \bar s_{\beta'} \gamma_\mu P_L d^\rho \bar s_{\rho'} \gamma^\mu P_L d^\beta = \bar s \gamma_\mu P_L d \bar s \gamma^\mu P_L d -  \bar s_\beta \gamma_\mu P_L d^\rho \bar s_\rho \gamma^\mu P_L d^\beta =0.
    \label{eq:gamma_LR}
    \end{equation}
  The vanished result is from the cancellation between the first and second term when the Fierz transformation is applied to the second term. Clearly, in the limit of $m_t \sim 0$, the box diagrams mediated by one $W$ and one ${\bf H}_3$ have no contributions to the $\Delta S=2$ process.  Hence, the nonvanished ${\cal H}^{WH_3}_{\Delta S=2}$ is from the $g^R_{31} g^{R*}_{32}$ term. In order to avoid the gauge dependence, we have to include the charged-Goldstone-boson contributions, where the dominant Yukawa coupling is $m_t V_{tq}/(\sqrt{2} m_W)\, \bar t_R q_L G^+$ (q=d,s).  In terms of the effective operators in Eq.~(\ref{eq:Qs}), we can write the effective Hamiltonian to be:
 \begin{equation}
 {\cal H}^{WH_3}_{\Delta S=2} = \frac{G^2_F V_{\rm CKM}}{16 \pi^2 } m^2_W \left( C^{LR}_{WH_{3}, 1} Q^{LR}_1 +  C^{LR}_{WH_{3}, 2} Q^{LR}_{2} \right)\,, \label{eq:W_H3_a}
 \end{equation}
 where the effective Wilson coefficients are given as:
 \begin{align}
 C^{LR}_{WH_3,1} &= 4 h^R_{21}  \left[y_W I^{WH_3}_{\rm Box}(y_W, y_t)-I^{GH_3}_{\rm Box}(y_W, y_t) \right]\,, ~ C^{LR}_{WH_{3}, 2} = 2 C^{LR}_{WH_3,1}\,, \nonumber \\
 I^{WH_3}_{\rm Box}(y_W, y_t) & = \frac{ y_t}{(y_t - y_W)^2} \left[ \frac{y_t -y_W}{1-y_t} + \frac{y_W \ln y_W}{1-y_W} + \frac{(y^2_t -y_W)\ln y_t}{(1-y_t)^2}\right]\,, \nonumber \\
 I^{GH_3}_{\rm Box} (y_W, y_t) & = -\frac{y^2_t}{2(1-y_t) (y_t -y_W)} - \frac{y_t y^2_W \ln y_W}{2(1-y_W)(y_t - y_W)^2} \nonumber \\
 & - \frac{y^2_t \left( y_t - (2- y_t) y_W\right) \ln y_t}{2(1-y_t)^2 (y_t -y_W)^2}\,.
 \end{align}
  With $m_{H_3}=1.5$ TeV,  the loop functions can be $ I^{WH_3}_{\rm Box}(y_W, y_t)\approx 0.68$ and $I^{G H_3}( y_W, y_t)\approx 0.02$. However, when $y_W$ factor is included, we obtain $y_W  I^{WH_3}_{\rm Box}(y_W, y_t)\approx 0.002$, which is smaller than $I^{G H_3}_{\rm Box}( y_W, y_t)$ by one order of magnitude; that is, $I^{GH_3}_{\rm Box}$ dominates. 
  
%  Due to $m_{W}\ll m_{H_3}$,  the loop function can be $ I^{WH_3}_{\rm Box}(y_W, y_t)\approx 0.68$ with $m_{H_3}=1.5$ TeV so it is %possible that the contribution of ${\cal H}^{WH_3}_{\Delta S=2}$ is significant and comparable with that mediated by the pure ${\bf H}%_3$ box diagrams.
 
% {\LARGE Estimate the size of this result and discuss the importance} 
 
 \subsubsection{\bf Box diagrams from the color-triplet diquark}
 
 The effective Hamiltonian through the mediation of  the diquark ${\bf H}_{3}$ shown in  Fig.~\ref{fig:box_W_H3}(b) can be written as:
 \begin{align}
 -i {\cal H}^{H_3}_{\Delta S=2} & = \frac{K_C}{2} \int \frac{d^4 q}{ (2 \pi )^4} \frac{{\cal N}_{H_3} }{(q^2 -m^2_W)^2(q^2 -m^2_t)^2 } \,,  \label{eq:H_H3}\\
  K_C & = (\bar K_b)_{\beta \alpha} (K^b)^{\alpha' \rho'} (\bar K_a)_{\beta' \alpha'} (K^a)^{\alpha \rho} = \frac{1}{4} \left( \delta^\rho_{\beta'} \delta ^{\rho'}_{\beta} + \delta^{\rho'}_{\beta'} \delta^\rho_\beta \right)\,, \nonumber \\
 {\cal N}_{H_3}  & = (\bar s_\rho \slashed{q} \chi^V_{21} d^\beta) ( \bar s_{\rho'} \slashed{q} \chi^V_{21} d^{\beta'} ) + m^2_t  (\bar s_\rho \chi^S_{21} d^\beta ) ( \bar s_{\rho'}  \chi^S_{21} d^{\beta'} )\,,  
 \end{align}
 where the crossed diagram by exchanging top-quark  and ${\bf H}_3$ is included, and the definitions of $\chi^V_{21}$ and $\chi^S_{21}$ can be found from  Eq.~(\ref{eq:chiVS}) in the appendix.  Using the Fierz transformations and the identities in Eq.~(\ref{eq:FT}), we find that the effective operators in Eq.~(\ref{eq:Qs}) can be all generated  from the box diagrams, and Eq.~(\ref{eq:H_H3}) can be formed as:
 \begin{align}
 {\cal H}^{H_3}_{\Delta S=2} & = \frac{G^2_F V_{\rm CKM} }{16 \pi^2 } m^2_W \left[ C^{VLL}_{H_3,1} Q^{VLL}_1 + C^{VRR}_{H_3,1} Q^{VRR}_1  + C^{LR}_{H_3,1} Q^{LR}_{1}  + C^{LR}_{H_3,2}Q^{LR}_2  \right.\nonumber \\
 & \left.  + C^{SLL}_{H_3,1} Q^{SLL}_1 + C^{SLL}_{H_3,2} Q^{SLL}_2 + C^{SRR}_{H_3,1} Q^{SRR}_1 + C^{SRR}_{H_3,2} Q^{SRR}_2\right]\,,
  \label{eq:H_H3_2}
 \end{align}
 where the associated effective Wilson coefficients at the $\mu=m_{H_3}$ scale are expressed as:
  \begin{align}
  C^{VLL}_{H_3,1}  & = 4 y_W I^{H_3}_{B1}(y_t) \left( h^L_{21}\right)^2 \,, ~ C^{VRR}_{H_3, 1} = 4 y_W I^{H_3}_{B1}(y_t) \left( h^R_{21}\right)^2\,, \nonumber  \\
   C^{LR}_{H_3, 1}  & = 4 y_W \left[ I^{H_3}_{B1}(y_t) + I^{H_3}_{B2}(y_t) \right] h^L_{21} h^R_{21}\,, ~ C^{LR}_{H_3,2} = -2 C^{LR}_{H_3,1} \,, \nonumber \\
   C^{SLL}_{H_3, 1} &= 2 y_W  I^{H_3}_{B2}(y_t) (h^L_{21})^2 \,, ~ C^{SLL}_{H_3, 2} = - \frac{C^{SLL}_{H_3, 1} }{4} \,, \nonumber \\
    C^{SRR}_{H_3, 1} &= 2 y_W  I^{H_3}_{B2}(y_t) (h^R_{21})^2 \,, ~ C^{SRR}_{H_3, 2} = - \frac{C^{SRR}_{H_3, 1} }{4}\,. \label{eq:Cchi_H3_i}
   \end{align}
   The loop functions $I^{H_3}_{B1}(y_t)$ and $I^{H_3}_{B2}(y_t)$ are defined as:
    \begin{align}
    I^{H_3}_{B1}(y) & = \frac{1+y}{2(1-y)^2} + \frac{y \ln y}{(1-y)^3}\,, \nonumber \\
    I^{H_3}_{B2}(y) & =- \frac{2 y}{(1 - y )^2} - \frac{y (1+y)\ln y}{(1-y)^3} \,.
    \end{align}
 From the interactions in Eq.~(\ref{eq:H_H3_2}), it can be seen that  eight different operators are involved. We will show that although the hadronic matrix elements of $Q^{VLL}_1$  and $Q^{VRR}_1$ are smaller than those of $Q^{SLL}_i$ and $Q^{SRR}_i$, due to $I^{H_3}_{B2}(y_t) \ll I^{H_3}_{B1}(y_t)$, their contributions indeed are comparable.

\section{$\epsilon'/\epsilon$ and $\epsilon_K$ with hadronic effects in the diquark model}

\subsection{Matrix elements for the $K\to \pi \pi$ decays}

The decay amplitudes for $K\to \pi \pi$ in terms of the isospin of $\pi\pi$ final state can be written as~\cite{Cirigliano:2011ny}:
  \begin{align}
  A(K^+ \to \pi^+ \pi^0 ) & = \frac{3}{2} A_2 e^{i \delta_2} \,, \nonumber \\
  A(K^0 \to \pi ^+ \pi^- ) & = A_0 e^{i \delta_0} + \sqrt{\frac{1}{2}} A_2 e^{i\delta_2}\,, \nonumber \\
  A(K^0 \to \pi^0 \pi^0 ) & = A_0 e^{i\delta_0} - \sqrt{2} A_2 e^{i\delta_2}\,
  \end{align}
  where $A_{0(2)}$ denotes the isospin $I=0 (2)$ amplitude; $\delta_{0(2)}$ is the strong phase and  $\delta_0 - \delta_2 = (47.5 \pm 0.9)^{\circ}$~\cite{Cirigliano:2011ny}.  The  experimental data indicate $Re A^{\rm exp}_0 = 27.04(1) \times 10^{-8}$ GeV and $Re A^{\rm exp}_2 =1.210(2) \times 10^{-8}$ GeV~\cite{PDG}. Using the isospin amplitudes, the direct CP violating parameter from new physics in $K$ system can be estimated by~\cite{Buras:2015yba}:
 \begin{equation}
 Re\left( \frac{\epsilon'}{\epsilon}\right) \approx - \frac{  \omega }{\sqrt{2} |\epsilon_K|} \left[ \frac{Im A_0}{ Re A_0} - \frac{Im A_2}{ Re A_2}\right]\,,  \label{eq:epsilon_p}
  \end{equation}
 where $\omega = Re A_2/Re A_0 \approx 1/22.35$ denotes the $\Delta I=1/2$ rule.
 From Eq.~(\ref{eq:epsilon_p}), it is seen that $\epsilon'/\epsilon$ is related to the ratios of hadronic matrix elements. In the  following, we summarize the relevant matrix elements for the involved operators that are from the tree-level and loop diagrams.

\subsubsection{ \bf $K\to \pi\pi$ hadronic matrix elements of the tree-level operators} 

Although  only one Feynman diagram is used to generate  the $\Delta S=1$ processes at the tree level, from Eq.~(\ref{eq:treeH}), twelve effective operators are involved in the processes, such as $Q_{1,2}$, $Q^{SLL,u}_{1-4}$ and their prime operators. 
The operators $Q_{1,2}$  are the same as those generated via the mediation of $W$-boson in the SM; thus, the associated hadronic matrix elements can be quoted from the SM calculations. However, the operators $Q^{SLL,u}_{i}$ are  new operators and do not mix with the SM operators; therefore, if  $Q_{1,2}$ and $Q^{SLL,u}$ are taken as two different classes of operators, we can  separately introduce their matrix elements.  According to the notations in~\cite{Buras:2015yba}, we thus define the new operators in terms of $Q_1$ and $Q_2$ as:
\begin{equation}
Q_{+}=\frac{1}{2} \left( Q_2 + Q_1 \right)\,, ~ Q_{-} = \frac{1}{2} \left( Q_2 - Q_1 \right)\,. 
\end{equation}
The isospin amplitudes for the $K\to \pi \pi$ decays in the SM can be given as~\cite{Buras:2015yba}: 
\begin{align}
ReA^{\rm SM}_0 &\approx  \frac{G_F V^*_{us} V_{ud}}{\sqrt{2}} z_{-} \langle Q_{-} \rangle_0 \left(1 + q_T \right)  \,, \nonumber \\
ReA^{\rm SM}_2 &\approx \frac{G_F V^*_{us} V_{ud}}{\sqrt{2}} z_{+} \langle Q_{+} \rangle_2 \,.
\end{align}
where $q_T= z_{+} \langle Q_{+} \rangle_0/( z_{-} \langle Q_{-} \rangle_0 )$,  $z_{\pm } =z_{2} \pm z_{1} $,  and the values of $z_{1,2}$ at $\mu=m_c$ are $z_1=-0.409$ and $z_2=1.212$~\cite{Buras:2015yba}. Because $q_T\lesssim 0.1$, we will ignore its contribution in the new physics study. In addition, we assume $ReA^{\rm SM}_{0(2)}\approx ReA^{\rm exp}_{0(2)}$ in the following analysis; that is, $\langle Q_{\pm} \rangle$ can be determined by the experimental data.

 Using the results obtained in~\cite{Aebischer:2018rrz}, the matrix elements arisen from the $Q^{SLL,u}_i$ operators for the isospin $I=0$ at the factorizable scale are given as:
 \begin{align}
 \langle Q^{SLL,u}_{1} \rangle_0 & = \frac{r^2(\mu)}{ 48} f_\pi\,, ~ \langle Q^{SLL,u}_{2} \rangle_0  = -\frac{r^2(\mu)}{ 24} f_\pi \,, \nonumber \\
  \langle Q^{SLL,u}_{3} \rangle_0 & = -\frac{r^2(\mu)}{ 4} f_\pi\,, ~ \langle Q^{SLL,u}_{4} \rangle_0  = 0\,,\label{eq:ME_tree_I0}
 \end{align}
with 
 \begin{equation}
r(\mu) =  \frac{ 2 m^2_K}{m_s(\mu) + m_d(\mu)} \,.
 \end{equation} 
The  matrix elements for the isospin $I=2$ are given as $\langle O^{SLL,u}_{i}\rangle_2 = \langle O^{SLL,u}_{i} \rangle_0/\sqrt{2}$.  Based on the DQCD approach, the matrix elements at the nonfactorizable scale $\Lambda$ can be expressed as~\cite{Aebischer:2018rrz}:
 \begin{align}
\langle Q^{SLL,u}_{1}(\Lambda) \rangle_I & = \left( 1 + \frac{4}{3} \hat{\Lambda}^2 \right)\langle Q^{SLL,u}_{1} \rangle_I + 4 \hat{\Lambda}^2 \langle Q^{SLL,u}_{2} \rangle_I\,, \nonumber \\
\langle Q^{SLL,u}_{2}(\Lambda) \rangle_I & = \left( 1 + \frac{4}{3} \hat{\Lambda}^2 \right)\langle Q^{SLL,u}_{2} \rangle_I + 2 \hat{\Lambda}^2 \langle Q^{SLL,u}_{1} \rangle_I - \frac{1}{2}  \langle Q^{SLL,u}_{3}\rangle_I\,, \nonumber \\
\langle Q^{SLL,u}_{3}(\Lambda) \rangle_I  & = \left( 1 + \frac{4}{3} \hat{\Lambda}^2 \right)\langle Q^{SLL,u}_{3} \rangle_I  -16 \hat{\Lambda}^2 \langle Q^{SLL,u}_{2} \rangle_I \,, \nonumber \\
\langle Q^{SLL,u}_{4}(\Lambda) \rangle_I  & =  \hat{\Lambda}^2 \left( -8 \langle Q^{SLL,u}_{1} \rangle_I + 2 \langle Q^{SLL,u}_{3} \rangle_I \right)\,,
 \end{align}
 where $\hat{\Lambda}$ is defined as:
  \begin{equation}
  \hat\Lambda = \frac{\Lambda}{4 \pi f_\pi} \left( 1 + \frac{m^2_\pi}{\Lambda^2_\chi} \right)\,, ~ \Lambda^2_\chi= \frac{m^2_K - f_K/f_\pi\, m^2_\pi}{f_K/f_\pi -1} \approx 1.15 \ \text{GeV}^2\,. \label{eq:Lambda_chi}
  \end{equation}
The matrix elements at a higher scale, e.g. $\mu > 1$ GeV, can be obtained through 
\begin{equation}
\langle Q^{SLL,u}_i (\mu) \rangle_{I} = \left( \delta_{ij} - \frac{\alpha_s}{4\pi}  \hat{\gamma}^{(0)}_{ij}  \ln\frac{\mu}{\mu_0} \right) \langle Q^{SLL,u}_i (\mu_0) \rangle_{I}\,, \label{eq:ME_mu}
\end{equation}
and the associated anomalous dimension matrix (ADM) in the basis of $(Q^{SLL,u}_{1}, Q^{SLL,u}_{2}, Q^{SLL,u}_3, Q^{SLL,u}_4)$ is~\cite{Aebischer:2018rrz}:
 \begin{align}
 \hat{\gamma}^{(0)SLL,u} = \left (
\begin{array}{cccc}
  6/N_c &  -6 & N_c/2 -1/N_c &  1/2 \\
0  & -6N_c + 6/N_c  &  1 & -1/N_c\\
-48/N_c + 24 N_c  & 24  & -2/N_c - 4 N_c  &  6\\
  48 & -48/N_c & 0 & 2N_c - 2/N_c
\end{array}
\right)\,, \label{eq:ga_SLLu}
  \end{align}
  with $N_c=3$. 
  According to the results in~\cite{Aebischer:2018rrz}, we show the numerical values of the $Q^{SLL,u}_i$ matrix elements for the $K\to \pi\pi$ decays  at $\mu=m_c=1.3$ GeV in Table~\ref{tab:QSLL_u}. We note that in terms of magnitude, the matrix elements of the prime operators  are the same as those of the unprimed operators, but they are  opposite in sign.
  
  \begin{table}[htp]
\caption{Value of hadronic matrix elements (MEs) in units of GeV$^3$ for $K\to \pi \pi$ from the $Q^{SLL,u}_i$ operators at the $\mu=1.3$ GeV.}
\begin{tabular}{c|cccc} \hline \hline 
 ME ~~ & $~~\langle Q^{SLL,u}_1 \rangle_I $~~ & ~~$\langle Q^{SLL,u}_2 \rangle_I  $~~ & ~~$\langle Q^{SLL,u}_3 \rangle_I $~~  & ~~$\langle Q^{SLL,u}_4 \rangle_I $~~ \\ \hline 
$I=0$ &  $-0.005$ & $-0.044$ & $-0.371$ & $-0.214$ \\ \hline 
$I=2$ &  $-0.003$ & $-0.031$ & $-0.262$ & $-0.151$ \\ \hline 
 
\end{tabular}
\label{tab:QSLL_u}
\end{table}%

To calculate the $K\to \pi \pi$ decay amplitudes, in addition to the hadronic matrix elements, we also need the effective Wilson coefficients at   $\mu=m_c$, which can be obtained using RG running from the $\mu=m_{H_3}$ scale. Therefore, for the operators $Q^{(\prime)SLL,u}_i$, the necessary ADM at the LO  QCD corrections can be found from Eq.~(\ref{eq:ga_SLLu}).  Since $Q_{1,2}$ mix with the  QCD and EW penguin operators, i.e. $Q_{3-10}$,  we basically need the $10\times 10$ ADM matrix for the operators $Q_{1-10}$. Since the mixture of $Q_{1,2}$ and $Q_{3-10}$ is dominated by the QCD penguin operators, we adopt the $6\times 6$ ADM for the new physics effects, and the ADM is
 given as~\cite{Buchalla:1995vs}:
 \begin{align}
 \hat{\gamma}^{(0)}_{QCD} = \left (
\begin{array}{cccccc}
  \frac{6}{N_c} &  6 & 0 &  0 & 0 & 0 \\
6  &  \frac{-6}{N_c} &  \frac{-2}{3N_c} & \frac{2}{3} & \frac{-2}{3N_c} & \frac{2}{3} \\
0   & 0  & \frac{-22}{3N_c}  &  \frac{22}{3} & \frac{-4}{3N_c} & \frac{4}{3} \\
0 & 0 & 6-\frac{2 f}{3N_c} & \frac{-6}{N_c} + \frac{2f}{3} & \frac{-2f}{3N_c} & \frac{2 f}{N_c}  \\
0 & 0 & 0 & 0 & \frac{6}{N_c} & -6  \\
0 & 0 & \frac{-2f}{3N_c} & \frac{2f}{3} & \frac{-3f}{3N_c} & \frac{-6(-1+N^2_c)}{N_c} + \frac{2f}{3}  
\end{array}
\right)\,, \label{eq:ADMQCD}
  \end{align}
with $f$ being the number of flavors. If we take the operators $Q_{1-6}$ as a basis,  from Eq.~(\ref{eq:treeH}), the corresponding Wilson coefficients can form a vector and be expressed as $C_T=(1, -1, 0, 0, 0, 0) \zeta^{LL}_{21}$ and $C'_T=(1,-1,0,0,0,0)\zeta^{RR}_{21}$ at the $m_{H_3}$ scale. Using RG evolution with ADM in Eq.~(\ref{eq:ADMQCD})~\cite{Buchalla:1995vs}, the Wilson coefficients at the $m_c$ scale can be obtained as:
 \begin{align}
 C_T(m_c) \approx (2.0,\, -2.0,\, 0,\, 0,\, 0,\, 0)\zeta^{LL}_{21}\,, 
 %
% C'_T(m_c) \approx (2.0,\, -2.0,\, 0,\,, 0,\, 0,\, 0.1)\zeta^{RR}_{21}
 \end{align} 
where we have ignored the effects that are less than or around $\pm 0.1$, and $C'_T(m_c)$ can be obtained from $C_T(m_c)$ using $\zeta^{RR}_{21}$ instead of $\zeta^{LL}_{21}$. 

Similarly, we can apply the same approach to the $Q^{(\prime)SLL,u}_{1-4}$ operators. From the Hamiltonian in Eq.~(\ref{eq:treeH}), the Wilson coefficients at the $\mu=m_{H_3}$ scale can be formed as  $C^{SLL,u}=(4,\, 4,\, 1,\, 1)\zeta^{LR}_{21}$ and $C'^{SLL,u}=(4,\, 4,\, 1,\, 1)\zeta^{RL}_{21}$. Using the ADM in Eq.~(\ref{eq:ga_SLLu}), the Wilson coefficients at $\mu=m_c$ can then be obtained as:
 \begin{equation}
 C^{SLL,u}(m_c)=(-5.44,\, 1.33,\, 2.41,\, 0.09)\zeta^{LR}_{21}\,.
 \end{equation}
We can obtain $C'^{SLL,u}(m_c)$ from $C^{SLL,u}(m_c)$ using  $\zeta^{RL}_{21}$ instead of $\zeta^{LR}_{21}$. 

 Following  Eqs.~(\ref{eq:treeH}) and (\ref{eq:epsilon_p}) and using the introduced matrix elements, the $Re(\epsilon'/\epsilon)$ from the tree-level diquark contributions can be formulated as:
  \begin{align}
   \left( \frac{\epsilon'}{\epsilon} \right)_T^{H_3} & = T^{(1/2)}_{H_3}  - T^{(3/2)}_{H_3} \,, \nonumber \\
 T^{(1/2)}_{H_3} & = \frac{2.0 r_1 y_W}{ z_{-}} Im\left[ \lambda_t \left( \zeta^{RR}_{21} - \zeta^{LL}_{21} \right) \right] \nonumber \\
 & - \frac{0.94  r_2 y_W }{2 Re A_0} Im \left[ \lambda_t \left( \zeta^{RL}_{21} - \zeta^{LR}_{21} \right) \right] \,, \nonumber \\
 T^{(3/2)}_{H_3} & = -  \frac{0.67 r_2 y_W }{2 Re A_2} Im\left[ \lambda_t  \left( \zeta^{RL}_{21} - \zeta^{LR}_{21} \right)\right]\,, \label{eq:ep_eK_T}
  \end{align}
%  \begin{align}
%  A^{H_3,T}_0 & \approx - \frac{G_F V^*_{ts} V_{td}}{\sqrt{2}} \frac{y_W}{2} \left[ 0.069 \left( \zeta^{RR}_{21}(m_c) - \zeta^{LL}_{21}(m_c) \right) 
  %[1+ c_1(\Lambda) +c_2(\Lambda) ] X_F \left( 1 - 4 \frac{\alpha_s}{4\pi} \ln m_c\right)
%  -0.195  \left( \zeta^{RL}_{21}(m_c) - \zeta^{LR}_{21}(m_c) \right) \right]\,, \nonumber \\
  %
 % A^{H_3,T}_2 & \approx - \frac{G_F V^*_{ts} V_{td}}{\sqrt{2}} \frac{y_W}{2} \left[ 
%  -0.137  \left( \zeta^{RL}_{21}(m_c)  - \zeta^{LR}_{21} (m_c) \right) \right]\,,
%  \end{align}
where the values of matrix elements in Table~\ref{tab:QSLL_u} have been  applied; the  $q_T$ related effect is neglected; $\lambda_t\equiv V^*_{ts} V_{td}$,
   \begin{equation}
  r_1 = \frac{\omega}{\sqrt{2} |\epsilon_K| V^*_{us} V_{ud}} \approx 64.76\,, ~ r_2 = \frac{G_F \omega}{2 |\epsilon_K|} \approx 1.17\times 10^{-4} \, {\rm GeV}^{-2}\,,
  \end{equation}
and $\zeta^\chi_{21}$  are determined  at the $\mu=m_{H_3}$ scale. Due to ${\cal H}_{\rm tree} \supset Q^{(\prime)}_1 - Q^{(\prime)}_2$ and $\langle Q_1\rangle_2 = \langle Q_2 \rangle_2$,   $T^{(3/2)}_{H_3}$ can only arise from the $Q^{SLL,u}_{i}$ operators.

\subsubsection{ \bf $K\to \pi \pi$ matrix elements of the QCD and EW penguin operators}

The  operators induced from the QCD and EW penguins for $\Delta S=1$ in the diquark model are similar to those generated in the left-right symmetric model~\cite{Bertolini:2012pu}, in which  the SM operators are included;  therefore, we can directly use the SM results for the $K\to \pi \pi$ decays. Using the Fierz transformations, it can be found that the operators $Q_{4,9, 10}$ can be expressed as: 
 \begin{align}
 Q_4 & = 2Q_{-} + Q_3 \,, ~Q_9 = \frac{3}{2} \left( Q_{+} - Q_{-} \right) - \frac{1}{2} Q_3 \,, \nonumber \\
  Q_{10} & = \frac{1}{2} \left( 3 Q_{+} + Q_{-} \right)- \frac{1}{2} Q_3\,. 
 \end{align}
Thus, the associated matrix elements can be written as:
 \begin{align}
 \langle Q_4 \rangle_0 & = 2\langle Q_{-} \rangle_0 + \langle Q_3\rangle_0\,, \langle Q_9 \rangle_0  = \frac{3}{2}\left(\langle Q_{+} \rangle_0 - \langle Q_{-} \rangle_0\right)- \frac{1}{2} \langle Q_3\rangle_0\,, \nonumber \\
 \langle Q_{10}\rangle_0 & = \frac{1}{2} \left( 3 \langle Q_{+}\rangle_0 + \langle Q_{-} \rangle_0 \right)- \frac{1}{2} \langle Q_3\rangle_0\,, ~ \langle Q_9 \rangle_2  =  \langle Q_{10} \rangle_2= \frac{3}{2}\langle Q_{+} \rangle_0\,, \label{eq:ME_penguin}
 \end{align}
 where $\langle Q_{-} \rangle_2=  \langle Q_{3} \rangle_2=0$ are applied.  From a native factorization, it can be found that $\langle Q_3 \rangle$  indeed is smaller than  $\langle Q_4 \rangle$ by a factor of $N_c$. If we drop the $\langle Q_{3} \rangle_0$ contributions, the matrix elements in Eq.~(\ref{eq:ME_penguin}) can be further simplified and are only related to $\langle Q_{\pm}\rangle$. It can be found that the same property  can be also applied to $\langle Q_5\rangle$ and $\langle Q_7 \rangle$; therefore, in the numerical estimates, we take the approximation by  neglecting  the $\langle Q_{3,5,7} \rangle$ effects. 
 
 The matrix elements for the $Q_{6,8}$ operators can be parametrized as~\cite{Buras:2015yba}:
 \begin{align}
 \langle Q_6 (\mu) \rangle_0 & =  -  (f_K - f_\pi) r^2(\mu) B^{(1/2)}_6\,, \nonumber \\
 \langle Q_8 (\mu) \rangle_0 & =  \frac{ f_\pi}{2}  r^2(\mu)  B^{(1/2)}_8 \,, \nonumber \\
 \langle Q_8 (\mu) \rangle_2  & = \frac{\sqrt{2} f_\pi}{4}  r^2(\mu) B^{(3/2)}_8\,,
 \end{align}
 where $B^{(1/2)}_{6,8}$ and $B^{(3/2)}_8$ are the nonperturbative parameters. 
% where $m_s(m_c) + m_d(m_c)\approx 0.114$ GeV, and the nonperturbative parameters are taken as $B^{1/2}_6 \approx 0.57$, $B^{(1/2)}_8 %\approx 1.0$, and $B^{(3/2)}_8\approx 0.76$.  
 We note that although the $Q^{(\prime)}_{8,10}$ operators do not appear in the Hamiltonian at the $\mu=m_{H_3}$ scale, they can be induced through RG evolution. 
 
 Moreover, the matrix elements of the prime operators can be obtained by reversing the signs of the unprimed operators. To summarize, from Eq.~(\ref{eq:epsilon_p}), we can formulate $Re(\epsilon'/\epsilon)$, which arises from the penguin diagrams in the diquark model, as:
 \begin{align}
 \left( \frac{\epsilon'}{\epsilon} \right)^{H_3}_{P} & = P^{(1/2)}_{H_3}  - P^{(3/2)}_{H_3} \,, \nonumber \\
 P^{(1/2)}_{H_3} & = a^{(1/2)}_{H_3 0} + a^{(1/2)}_{H_3 6} B^{(1/2)}_6\,, \nonumber \\
 P^{(3/2)}_{H_3} & = a^{(3/2)}_{H_3 0} + a^{(3/2)}_{H_3 8} B^{(3/2)}_8\,, \label{eq:ep_eK_P}
  \end{align}
 where $a^{(1/2)}_i$ and $a^{(3/2)}_i$ are given by:
 \begin{align}
 a^{(1/2)}_{H_3 0} & \approx  \frac{ r_{1}}{2 z_{-}} Im\left[ \lambda_t \left( 4 \Delta y^{H_3}_4(m_c) -3 \Delta y^{H_3}_9 (m_c)+ \Delta y^{H_3}_{10} (m_c)\right) \right]  \nonumber \\
 & +   \frac{r_2 \langle Q_8 \rangle_0}{Re A_0}Im\left[ \lambda_t \Delta y^{H_3}_8(m_c) \right] \,, \nonumber \\
a^{(1/2)}_{H_3 6} & \approx \frac{r_2 \langle Q_6\rangle_0 }{B^{(1/2)}_{6} ReA_0}  Im\left[ \lambda_t \Delta y^{H_3}_6(m_c) \right]\,, \nonumber \\
a^{(3/2)}_{H_3 0} & \approx  \frac{ 3 r_{1}}{2 z_{+}} Im\left[ \lambda_t \left(  \Delta y^{H_3}_9(m_c) + \Delta y^{H_3}_{10}(m_c) \right) \right] \,, \nonumber \\
a^{(3/2)}_{H_3 8} & \approx \frac{r_2 \langle Q_8 \rangle_2}{B^{(3/2)}_8 ReA_2}  Im\left[ \lambda_t \Delta y^{H_3}_8 (m_c)\right]\,, \label{eq:a_P}
 \end{align}
 with $\Delta y^{H_3}_i (m_c) = y^{H_3}_i(m_c)  - y'^{H_3}_{i}(m_c)$.   Using the leading order $10\times 10$  ADM for the $Q_{1-10}$ operators~\cite{Buchalla:1995vs}, the effective Wilson coefficients appearing in Eq.~(\ref{eq:a_P}) at $\mu=m_c$ can be obtained as:
  \begin{align}
  \Delta y^{H_3}_4 (m_c) & \approx  -0.70 \delta y^{H_3}_3 + 1.09 \delta y^{H_3}_4 - 0.10 \delta y^{H_3}_5 -0.56 \delta y^{H_3}_6\,, \nonumber \\
    \Delta y^{H_3}_6 (m_c) & \approx  -0.10 \delta y^{H_3}_3 -0.47 \delta y^{H_3}_4 + 0.93 \delta y^{H_3}_5 + 3.18 \delta y^{H_3}_6 + 0.12 \delta y^{H_3}_9\,, \nonumber \\
    \Delta y^{H_3}_8(m_c) & \approx 1.07 \delta y^{H_3}_7\,, \nonumber \\
    \Delta y^{H_3}_9(m_c) & \approx 1.36 \delta y^{H_3}_9 \,, \nonumber \\
     \Delta y^{H_3}_{10}(m_c) & \approx -0.65 \delta y^{H_3}_9 \,, \label{eq:Delta_yH3}
  \end{align}
 where we have dropped the operator mixing effects that are smaller than $10\%$, and $\delta y^{H_3}_i = y^{H_3}_i - y'^{H_3}_i$ denote the quantities at the $\mu=m_{H_3}$ scale.  From  Eq.~(\ref{eq:a_P}), it can be seen that the involved hadronic effects explicitly shown in $Re(\epsilon'/\epsilon)^P_{H_3}$ now are  only  $\langle Q_{6,8}\rangle$. 

\subsubsection{\bf $K\to \pi\pi$ matrix element of the CMOs}

To estimate the $K\to \pi \pi$ hadronic matrix element via the operators $Q^{(\prime)}_{8G}$, we take the results obtained by a DQCD approach as~\cite{Buras:2018evv}:
 \begin{equation}
 \langle \pi \pi | C^{-}_{8G} Q_{8G}(-) | K \rangle \approx C^{-}_{8G}(\mu)   \frac{9}{11} \frac{m^2_\pi}{\Lambda^2_\chi} \frac{m^2_K f_\pi}{m_s(\mu) + m_d(\mu)}\,, %\approx (4.1 \times 10^{-3} \ {\rm GeV}^2) \, C^-_{8G}(\mu)\,,
 \end{equation}
where $Q_{8G}(-)\equiv g_s/(16\pi^2)\bar s \sigma^{\mu \nu} T^a \gamma_5 d G^a_{\mu \nu}$, $C^-_{8G}(\mu)$ is the effective Wilson coefficient with mass dimension $(-1)$ at the $\mu$ scale, and  $\Lambda_{\chi}$ can be found in Eq.~(\ref{eq:Lambda_chi}).
  Thus, the Kaon direct CP violation arisen from CMOs can be simply estimated as:
\begin{align}
 Re\left( \frac{\epsilon'}{\epsilon}\right)_{8G}  & \approx - \frac{\omega}{\sqrt{2} |\epsilon_K|} \frac{(Im A_0)_{8G}}{Re A_0} \nonumber \\
 & \approx  - (4.1 \times 10^{-3} \ {\rm GeV}^2) \frac{\omega}{\sqrt{2} |\epsilon_K| ReA_0} Im( C^-_{8G}(m_c))  \,.  \label{eq:epsilon_p1} 
 %= - \frac{ a \omega }{\sqrt{2} |\epsilon_K|} \left[ \frac{Im A_0}{ Re A_0} (1 - \hat\Omega_{\rm eff}) - \frac{1}{a} \frac{Im A_2}{ Re 
 %A_2}\right]\,,  \nonumber \\
 %C^-_{8G}(\mu) & = \frac{ G_F}{ \sqrt{2}} V^*_{ts} V_{td} \left(m_d C'^{H_3}_{8G} (\mu) - m_s C^{H_3}_{8G} (\mu)\right) \nonumber \\
% & =  \frac{ G_F}{ \sqrt{2}} V^*_{ts} V_{td} m_t y_W I_{G2}(y_t) \eta_{8G}\left( h^R_{21} - h^L_{21} \right)\,, 
  \end{align}
%where the definitions of $C^{(\prime)}_{8G}$ shown in Eq.~(\ref{eq:CH3_8G}) are applied to the second line of Eq.~(\ref{eq:epsilon_p2}), and $%\eta_{8G}=0.418$ is the RG evolution factor from $m_{H_3}=1.5$ TeV to $m_c=1.3$ GeV.
 With $|\epsilon_K|= 2.228\times 10^{-3}$ and $ReA_0 = 27.04 \times 10^{-8}$ GeV, Eq.~(\ref{eq:epsilon_p1}) can be expressed as:
 \begin{equation}
 Re\left( \frac{\epsilon'}{\epsilon}\right)_{8G}   \approx  - (1.74 \times 10^{5} \ {\rm GeV}) \times Im( C^-_{8G}(m_c))  \,.  \label{eq:epsilon_p3}
 \end{equation} 
 
 According to the Hamiltonian shown in Eq.~(\ref{eq:CMD}), we can write the $C^{H_3 -}_{8G}$ in the diquark model at $\mu=m_c$ as:
  \begin{align}
  C^{H_3-}_{8G}(m_c) & = - \frac{ G_F}{ \sqrt{2}} V^*_{ts} V_{td}  \eta_{8G} \left(m_d C'^{H_3}_{8G}  - m_s C^{H_3}_{8G} \right) \nonumber \\
& \approx   - \frac{ G_F}{ \sqrt{2}} V^*_{ts} V_{td} m_t y_W I_{G2}(y_t) \eta_{8G}\left( h^R_{21} - h^L_{21} \right)\,, \label{eq:Cm_8G}
 \end{align}
 where the definitions of $C^{(\prime)}_{8G}$ shown in Eq.~(\ref{eq:CH3_8G}) are applied to the second line, $g^R_{32}/g^L_{32}\approx  1$ is used, and  $\eta_{8G}\approx 0.418$ is the RG evolution factor from $m_{H_3}=1.5$ TeV to $m_c=1.3$ GeV. For the study of new physics effects, we only consider the leading-order QCD ADM for the operators $Q_{1-6}$, $O_{7\gamma}$, and $Q_{8G}$~\cite{Buchalla:1995vs}.

 \subsection{ $\Delta S=2$ in the diquark model}
 
 Using the effective Hamiltonian in  Eq.~(\ref{eq:H_DS2}), the hadronic  matrix element of $K^0$-$\bar K^0$ mixing is written as:
\begin{equation}
 M^{*}_{12} = \langle \bar K^0 | {\cal H}_{\Delta S=2} | K^0 \rangle\,. \label{eq:M12}
 \end{equation}
Accordingly, the $K$-meson mixing parameter and indirect CP violating parameter can be obtained as:
\begin{equation}
\Delta M_K \approx  2 Re M_{12} \,,  \quad
\epsilon_K \approx \frac{e^{i \pi/4}}{\sqrt{2} \Delta M^{\rm exp}_K} Im M_{12}\,, \label{eq:DMeK}
\end{equation}
where  the small contribution of $Im A_0/ReA_0$ from $K\to \pi \pi$ in  $\epsilon_K$ has been neglected. Since $\Delta M_K$ is  measured well, we  directly take the $\Delta M_K$ data for the denominator of $\epsilon_K$. It has been found that  the short-distance SM result on $\Delta M_K$ can explain the  data by $\sim 70\%$, and the long-distance effects may contribute another~$20-30\%$ with a large degree of uncertainty~\cite{Buras:2014maa}. Conservatively, the new physics  can have the contribution with a $20\%$ of  the experimental value. Hence, to investigate the new physics contributions to $\Delta M_K$ and $\epsilon_K$, we use the formalism obtained in~\cite{Buras:2001ra}, which is  given as:
 \begin{align}
 \langle\bar K^0 | {\cal H}_{\Delta S=2} | K^0 \rangle & = \frac{G^2_F V_{\rm CKM} }{48 \pi^2} m^2_W m_K f^2_K  \left\{ P^{VLL}_1 \left[ C^{VLL}_{1}(\mu_t) + C^{VRR}_{1}(\mu_t)\right] \right. \nonumber \\
 & + P^{LR}_1 C^{LR}_{1} (\mu_t) + P^{LR}_2 C^{LR}_{2} (\mu_t)+ P^{SLL}_{1}\left[ C^{SLL}_{1}(\mu_t) + C^{SRR}_{1}(\mu_t)\right] \nonumber \\
 & \left. + P^{SLL}_2 \left[ C^{SLL}_{2}(\mu_t) + C^{SRR}_{2}(\mu_t)\right]\right\}\,, \label{eq:KK}
 \end{align}
where the Wilson coefficients $C^{\chi}_{i}$ are taken at the $\mu_t=m_t$ scale, and the values of $P^\chi_i$ at $\mu=2$ GeV are shown as:
 \begin{align}
 P^{VLL}_1 & \approx 0.48\,, \quad P^{LR}_1\approx  -36.1\,, \quad  P^{LR}_2 \approx59.3\,, \nonumber \\
 P^{SLL}_1 & \approx -18.1\,, \quad P^{SLL}_2\approx 32.2 \,. 
 \end{align}
Since the Wilson coefficients $C^{\chi}_{H_3,i}$ in the diquark model are obtained at $\mu=m_{H_3}$, due to $m_t < m_{H_3}$, we have to use the RG evolution to get $C^{\chi}_{H_3,i}(\mu_t)$.   For comparison, we separate  the discussions of Fig.~\ref{fig:box_W_H3}(a) and (b) in the following analysis.

  \subsubsection{ \bf Box diagrams with one $W$ and one ${\bf H}_3$} 
  
  According to Eq.~(\ref{eq:W_H3_a}), the related operators arisen from Fig.~\ref{fig:box_W_H3}(a) are $Q^{LR}_1$ and $Q^{LR}_2$, and the associated Wilson coefficients are $C^{LR}_{H_3,1}$ and $C^{LR}_{H_3,2}$. To obtain the $C^{LR}_{H_3,1(2)}$ at the $\mu_t$ scale, we adopt the leading  QCD corrections, where the one-loop ADM for $(Q^{LR}_1, Q^{LR}_2)$ is given as~\cite{Buras:2001ra}:
 \begin{align}
 \hat{\gamma}^{(0)LR} = \left (
\begin{array}{cc}
 2  &  12  \\
0  & -16  \\
  \end{array}
\right)\,.
  \end{align}
Using the ADM, we can obtain the $C^{LR}_{H_3,i}(\mu_t)$ as:
 \begin{align}
 C^{LR}_{WH_3,1} (\mu_t) & = \eta^{3/21} C^{LR}_{WH_3,1} \,, \nonumber \\
 C^{LR}_{WH_3,2} (\mu_t) & = \frac{2}{3} \left( \eta^{3/21}  - \eta^{-24/21} \right) C^{LR}_{WH_3,1} + \eta^{-24/21} C^{LR}_{WH_3,2} \,, 
 \end{align}
  with $\eta=\alpha^{(6)}_s(m_{H_3})/\alpha^{(6)}_s(m_t)$. Using the result of $C^{LR}_{WH_3, 2}=2  C^{LR}_{WH_3, 1}$, the $K^0-\bar K^0$ mixing matrix element is expressed as:
  \begin{align}
  \langle \bar K^0 |  {\cal H}^{WH_3}_{\Delta S=2} | K^0 \rangle & = \frac{G^2_F V_{\rm CKM}}{48 \pi^2 } m^2_W m_K f^2_K  \left(\eta^{3/21} P^{LR}_1 \right. \nonumber \\
  & \left. + \frac{2}{3} \left( \eta^{3/21} +2 \eta^{-24/21} \right) P^{LR}_2  \right) C^{LR}_{WH_3,1}  \label{eq:M12_WH3}\,.
 %+ \eta^{-24/21} P^{LR}_2 C^{LR}_{WH_3, 2} \right]\,.
  \end{align}
     
  \subsubsection{\bf Box diagrams with two ${\bf H}_3$ }
 
 The situation for Fig.~\ref{fig:box_W_H3}(b) is more complicated. From Eq.~(\ref{eq:H_H3_2}), it can be seen that $\langle \bar K^0  |{\cal H}^{H_3}_{\Delta S=2} | K^0 \rangle $  involve five  hadronic effects, i.e., $P^{VLL}_1$, $P^{LR}_{1,2}$, and $P^{SLL}_{1,2}$. Although the magnitude of $P^{VLL}_1$ is much smaller than that of $|P^{SLL}_{1(2)}|$, when including the loop functions with $I_{B2}\ll I_{B1}$,  $I_{B1} P^{VLL}_1$ and $I_{B2} P^{SLL}_{1(2)}$ become comparable. In addition, although the magnitudes of $P^{LR}_{1,2}$ are larger than the others and the associated loop function is $I_{B1}$, because the Yukawa couplings are $h^{L}_{21} h^{R}_{21}$,  either of them might be small.  Hence, we should retain  all contributions at the moment.

To estimate the Wilson coefficients at $\mu_t$, in addition to the ADM shown in Eq.~(\ref{eq:gamma_LR}), we  need the ADMs for $Q^{VLL}_1$ and $Q^{SLL}_{1,2}$, where they are given as~\cite{Buras:2001ra}:
 \begin{align}
  \hat{\gamma}^{(0)VLL} =4\,,~
   \hat{\gamma}^{(0)SLL} = \left (
\begin{array}{cc}
 -10  &  1/6  \\
-40 & 34/3  \\
  \end{array}
\right)\,.
   \end{align} 
 Using $C^{LR}_{H_3,2}=-2C^{LR}_{H_3,1}$ and $C^{SLL}_{H_3,2} = -C^{SLL}_{H_3,1}/4$, the Wilson coefficients at $\mu_t$ can then be expressed as:
   \begin{align}
   C^{SLL}_{H_3, 1}(\mu_t) &= \eta^{6/21} C^{VLL}_{H_3,1}\,, \nonumber \\
   C^{LR}_{H_3,1} (\mu_t) & = \eta^{3/21} C^{LR}_{H_3,1} \,, \nonumber \\
 C^{LR}_{H_3,2} (\mu_t) & = \frac{2}{3} \left( \eta^{3/21}  - 4 \eta^{-24/21} \right) C^{LR}_{H_3,1}  \,, \nonumber \\
 %
%    C^{SLL}_{H_3,1}(\mu_t) & = \left( \frac{8\left(\eta^{r_2 } - \eta^{r_1} \right)}{\sqrt{241}}  + \frac{1}{2} \left( \eta^{r_2} + \eta^{r_1} \right) \right) %C^{SLL}_{H_3,1} + \frac{30\left( \eta^{r_2} - \eta^{r_1}\right)}{\sqrt{241}}  C^{SLL}_{H_3,2} \,, \nonumber \\
C^{SLL}_{H_3,1}(\mu_t) & =  \left( \frac{\eta^{r_2 } - \eta^{r_1} }{2\sqrt{241}}  + \frac{1}{2} \left( \eta^{r_2} + \eta^{r_1} \right) \right)  C^{SLL}_{H_3,1} \,, \nonumber \\
   %
 %  C^{SLL}_{H_3,2}(\mu_t) & = -\frac{\left(\eta^{r_2 } - \eta^{r_1} \right)}{8\sqrt{241}} C^{SLL}_{H_3,1} +\left( - \frac{8\left(\eta^{r_2 } - \eta^{r_1} %\right)}{\sqrt{241}}  + \frac{1}{2} \left( \eta^{r_2} + \eta^{r_1} \right) \right) C^{SLL}_{H_3,2} \,, 
 C^{SLL}_{H_3,2}(\mu_t) & =\left(  \frac{15\left(\eta^{r_2 } - \eta^{r_1} \right)}{8\sqrt{241}}  - \frac{1}{8} \left( \eta^{r_2} + \eta^{r_1} \right) \right) C^{SLL}_{H_3,1} \,, 
 \label{eq:C_mut}
   \end{align} 
   with $r_1=(\sqrt{241}+1)/21$ and $r_2= - (\sqrt{241} -1)/21$. Since QCD does not distinguish chirality, Eq.~(\ref{eq:C_mut}) can be directly applied to $C^{VRR}_1(\mu_t)$ and $C^{SRR}_{i}(\mu_t)$. 
 
\section{Constraints from the $\Delta S=2$ process }

\subsection{ Experimental and theoretical inputs} 

For the numerical analysis, in addition to the values of theoretical parameters, in this section, we introduce the experimental data used to bound the free parameters. The data of the $\Delta S=2$ process are given as~\cite{PDG}:
   \begin{align}
  \Delta M^{\rm exp}_K & =(3.482\pm 0.006) \times 10^{-15} \text{ GeV} \,, ~  \epsilon^{\rm exp}_K =  (2.228\pm 0.011)\times 10^{-3} \,. 
  \end{align}
Since $\epsilon_K$ in the SM fits well with the experimental data~\cite{Buchalla:1995vs}, we use
 \begin{equation}
 \left|\epsilon^{\rm NP}_K \right| \leq  0.4 \times 10^{-3}  \label{eq:eNP_K}
 \end{equation} 
 to bound the new physics effects~\cite{Buras:2015jaq}.  The uncertainties of the NLO~\cite{Herrlich:1993yv} and NNLO~\cite{Brod:2011ty} QCD corrections to the short-distance contribution to $\Delta M_K$ in the SM are somewhat large, so we take the combination of  the short-distance (SD) and long-distance (LD) effects as $\Delta M^{\rm SM}_K (SD + LD)    = (0.80 \pm 0.10) \Delta M^{\rm exp}_K$~\cite{Buras:2014maa}. Thus, the new physics contribution to $\Delta M_K$ is required to satisfy: 
  \begin{equation}
   |\Delta M^{\rm NP}_K|\leq  0.2\, \Delta M^{\rm exp}_K  \,. \label{eq:DMNP_K}
  \end{equation}

With the Wolfenstein parametrization~\cite{Wolfenstein:1983yz},  the CKM matrix elements can be taken as:
 \begin{align}
 V_{ud} & \approx V_{cs} \approx 1-\lambda^2/2\,, \ V_{us}\approx - V_{cd}\approx \lambda = 0.225\,, ~V_{ub}\approx 0.0038 e^{-i\phi_3}\,, \ \phi_3=73.5^{\circ}\,, \nonumber\\
 V_{cb} & \approx -V_{ts} \approx  0.0407\,, ~V_{td} \approx 0.0088 e^{-i\phi_2}\,, ~ \phi_2 \approx 23.4^{\circ} \label{eq:CKM}
 \end{align}
where $V_{cb}$ and $V_{ub}$ are taken from the averages of inclusive and exclusive semileptonic decays~\cite{Buras:2015qea}; the $\phi_3$ angle is the central value averaged by the heavy flavor averaging group (HFLAV) through all  charmful two-body $B$-meson decays~\cite{Amhis:2016xyh}, and $\phi_2$ is determined through  the inputs of Eq.~(\ref{eq:CKM}). The particle masses used to estimate the numerical values are given as:
  \begin{align}
  & m_W \approx 80.385 \text{ GeV} \,, ~m_t \approx 165 \text{ GeV}\,, ~m_K\approx 0.489 \text{ GeV}\,,\nonumber \\
   & m_c\approx 1.3 \text{ GeV}\,, ~m_s(m_c) \approx  0.109 \text{ GeV}\,, ~m_d (m_c) \approx 5.44 \text{ MeV}\,.
  \end{align}

\subsection{ $\Delta M_K$ and $\epsilon_K$ from ${\cal H}^{WH_3}_{\Delta S=2}$ }

The involved parameters for the $\Delta S=2$ process in the diquark model contain  $g^L_{31,32}$, $g^R_{31,32}$, and $m_{H_3}$.  However, it was found that the new parameters $h^{L,R}_{21}$, defined in Eq.~(\ref{eq:hLR_21}),  are more useful to study the diquark effects for the $\epsilon_K$ and $\epsilon'/\epsilon$. Generally, the CP phases of $g^{L,R}_{31,32}$ are free variables; in order to simplify the numerical analysis, we assume that their CP phases are the same as $V^*_{ts} V_{td}$ although this assumption is not necessary. That is, we will take $h^{L,R}_{21}$ to be real parameters, and the CP violating source is uniquely dictated by the KM phase. In sum, there are three new free parameters for the $\Delta S=2$ process in this study is three, which  are $h^{L,R}_{21}$ and $m_{H_3}$. 

\begin{figure}[phtb]
\includegraphics[scale=0.60]{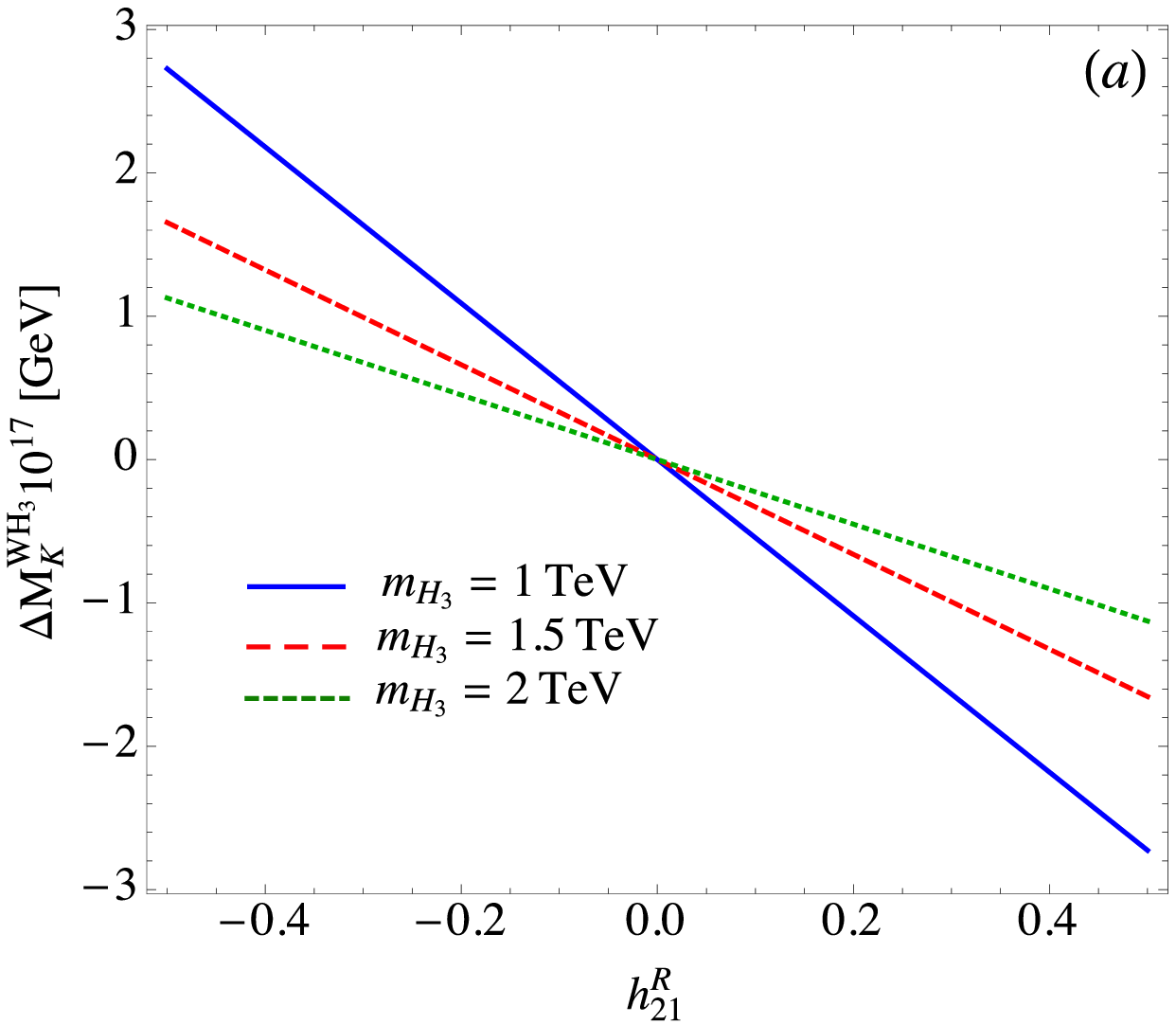}
\includegraphics[scale=0.60]{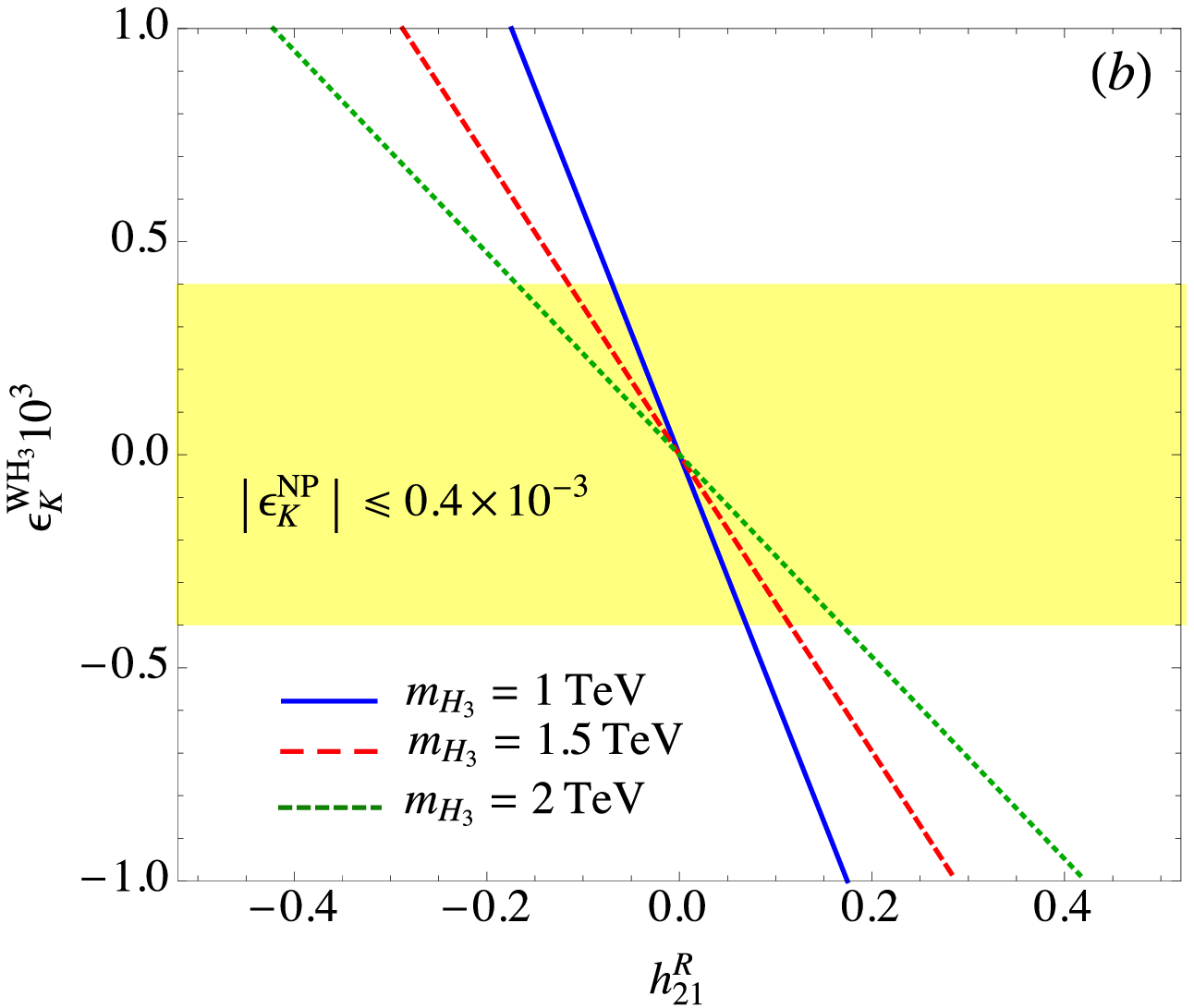}
 \caption{ Plots, which are from the $W-{\bf H_3}$ box diagrams, for (a) $\Delta M_K$ (in units of $10^{-17}$) and (b) $\epsilon_K$ (in units of $10^{-3}$) as a function of $h^R_{21}$, where the solid, dashed, and dotted lines represent the contributions of $m_{H_3}=(1,\, 1.5,\, 2)$ TeV, respectively. The band denotes the required limit shown in Eq.~(\ref{eq:eNP_K}). }
\label{fig:Box_WH3_ab}
\end{figure}

 Since ${\cal H}^{WH_3}_{\Delta S=2}$ only  depends on $h^R_{21}$ and $m_{H_3}$,  we can use the $\Delta S=2$ process to directly bound these parameters. Therefore, based on the transition matrix elements given in Eq.~(\ref{eq:M12_WH3}), we plot $\Delta M^{WH_3}_{K}$ (in units of $10^{-17})$ and $\epsilon^{WH_3}_K$ (in units of $10^{-3}$) as a function of $h^R_{21}$ in Fig.~(\ref{fig:Box_WH3_ab}), where the solid, dashed, and dotted lines represent the contributions of $m_{H_3}=(1,\, 1.5,\, 2)$ TeV, respectively. From the results, it can be clearly seen that the mass difference between $K_L$ and $K_S$, which arise from the $W-{\bf H}_3$ box diagrams, is far smaller than the required limit of $|\Delta M^{\rm NP}_{K}| \leq 0.2 \Delta M^{\rm exp}_{K}$ shown in Eq.~(\ref{eq:DMNP_K}). Since $\Delta M^{WH_3}_{K}$ and $\epsilon^{WH_3}_K$ originate from the same box diagrams, due to the CP phase of $V^*_{ts} V_{td}$ being of ${\cal O}(1)$, it can be expected that $\epsilon_K$ of  ${\cal O}(10^{-3})$ can constrain  the free parameters to a greater degree. The situation can be confirmed from Fig.~(\ref{fig:Box_WH3_ab})(b), where the range of $h^R_{21}$ is limited when the required limit of $|\epsilon^{\rm NP}_{K}|\leq 0.4\times 10^{-3}$ is imposed. For instance, using $m_{H_3}=1.5$ TeV, we obtain $|h^R_{21}|\lesssim  0.11$. 

  %%%%

\subsection{ $\Delta M_K$ and $\epsilon_K$ from  ${\cal H}^{H_3}_{\Delta S=2}$ }

As discussed before,  eight effective operators are involved in the purely ${\bf H}_{3}$-mediated box diagrams for the $\Delta S=2$ process. Since the hadronic effects have the properties of $P^{VLL(VRR)}_1 \ll |P^{SLL(SRR)}_{1,2}|$, the contributions from  $Q^{VLL(VRR)}_{1}$ are comparable to those from $Q^{SLL(SRR)}_{1,2}$ due to  the associated loop functions in the former and latter satisfying $I^{H_3}_{B1}(y_t) \gg I^{H_3}_{B2}(y_t)$. In addition, it can be seen from Eq.~(\ref{eq:Cchi_H3_i}) that the Wilson coefficients $C^{VLL(RLL)}_1$ and $C^{SLL(SRR)}_{1,2}$ depend on $h^{L(R)}_{21}$ in quadratic form. Therefore, it is of interest to understand their contributions to $\Delta M_K$ and $\epsilon_K$ without the $C^{LR}_{1, 2}$ effects, where  $C^{LR}_{1, 2}\propto h^L_{21}h^R_{21}$ and the associated loop functions show up in the form of $I^{H_3}_{B1}(y_t)+I^{H_3}_{B2}(y_t)$. Thus, taking $m_{H_3}=1.5$ TeV, $h^L_{21}=0$, and $h^R_{21}= 0.11$, where the chosen values obey the bound from $\epsilon^{WH_3}_K$, we find:
 \begin{equation}
 \Delta M^{H_3}_K \approx - 2.75\times 10^{-23}~\text{GeV} \,, ~ \epsilon^{H_3}_K \approx -2.90 \times 10^{-9}\,.
 \end{equation}
 Clearly, the contributions from the $Q^{VLL(VRR)}_{1}$ and $Q^{SLL(SRR)}_{1,2}$ operators that are induced from the ${\bf H}_{3}$ box diagrams are small and negligible. Since the behavior of $h^L_{21}$ is the same as that of $h^R_{21}$, the conclusion will not  change even with  $h^{L}_{21}\sim {\cal O}(10)$, with the exception of $h^L_{21}\sim {\cal O}(100)$. In addition, it is not necessary to combine ${\cal H}^{WH_3}_{\Delta S=2}$ and ${\cal H}^{H_3}_{\Delta S=2}$  because  the pure $h^R_{21}$ effect in ${\cal H}^{H_3}_{\Delta S=2}$ as shown above cannot compete with that in ${\cal H}^{WH_3}_{\Delta S=2}$.

\begin{figure}[phtb]
\includegraphics[scale=0.60]{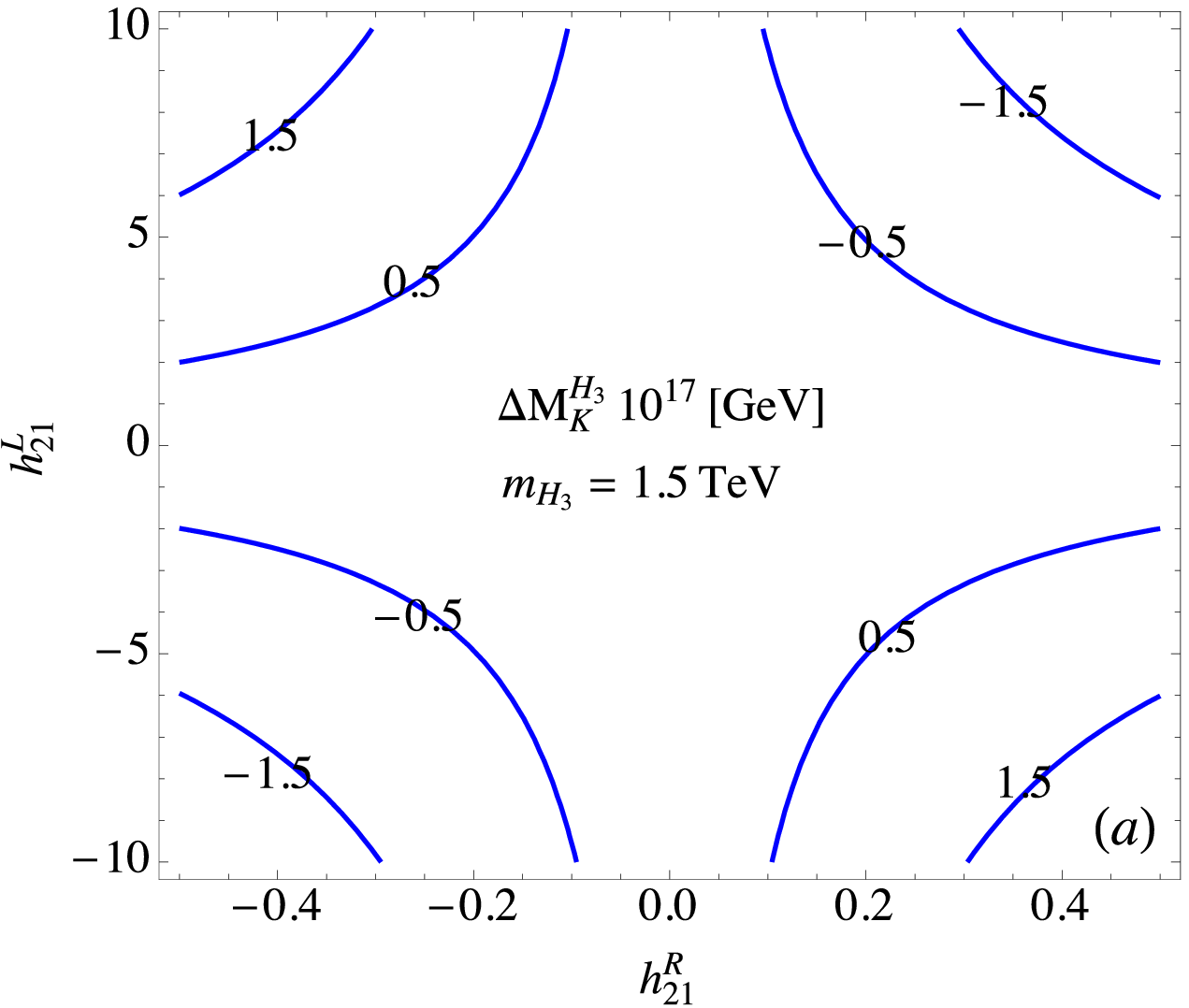}
\includegraphics[scale=0.60]{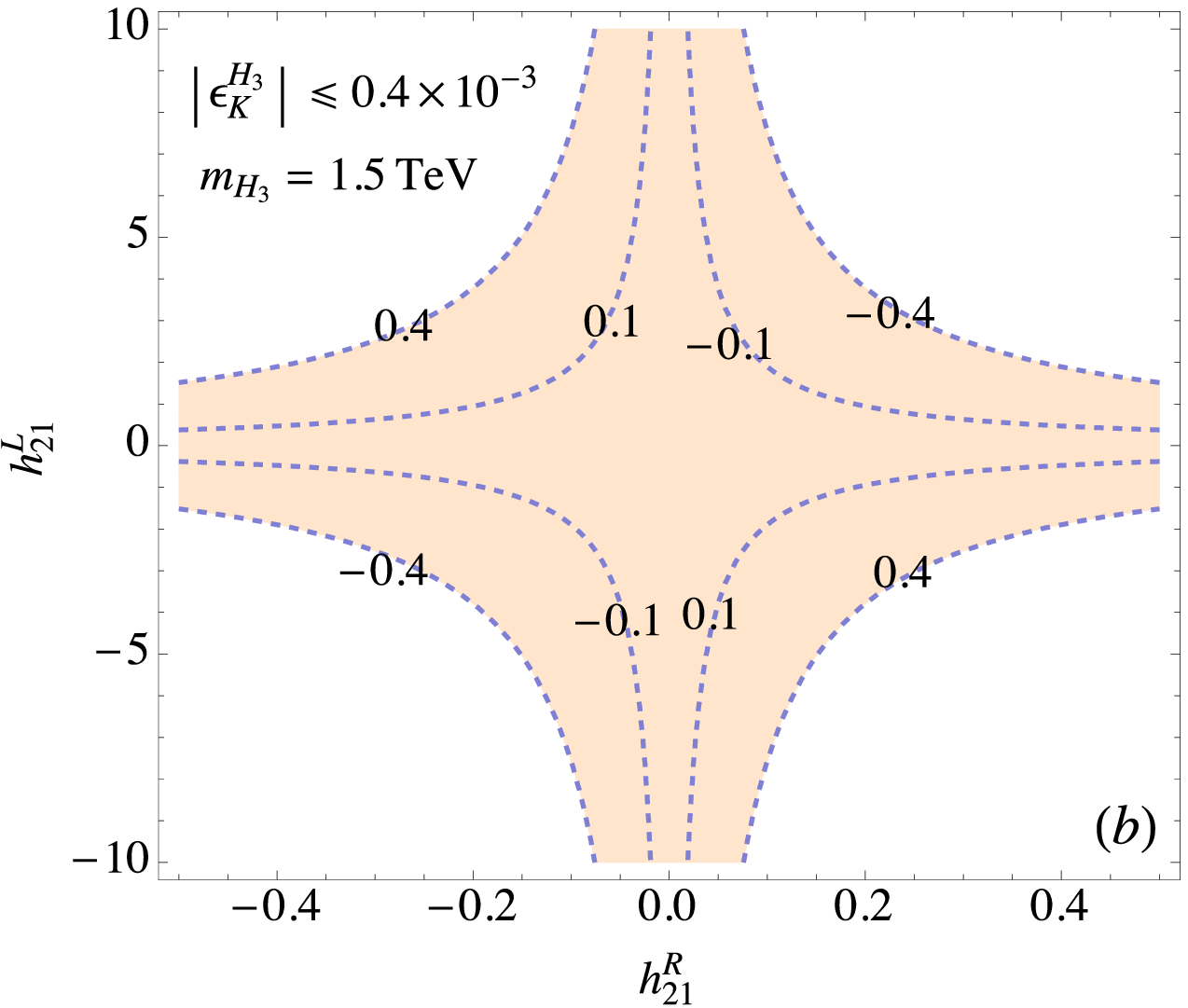}
 \caption{ Contours, which arise from the ${\bf H}_3-{\bf H}_3$ box diagrams, for (a) $\Delta M_K$ (in units of $10^{-17}$) and (b) $\epsilon_K$ (in units of $10^{-3}$) as a function of $h^L_{21}$ and $h^R_{21}$, where $m_{H_3}=1.5$ TeV is used. }
\label{fig:Box_H3_ab}
\end{figure}

 The ${\bf H}_{3}$ box diagrams could play an important role through the $C^{LR}_{1,2}$ effects. In addition to the loop function $I^{H_3}_{B1}(y_t)$, the enhancement factors are from  the associated hadronic effects $|P^{LR}_{1,2}|$, which are larger than the others.  For clarity, we make contour plots for $\Delta M^{H_3}_{K}$ (in units of $10^{-17}$) and $\epsilon^{H_3}_K$ (in unit of $10^{-3}$) as a function of $h^L_{21}$ and $h^{R}_{21}$ in Fig.~\ref{fig:Box_H3_ab}, where we fix $m_{H_3}=1.5$ TeV. From the plots, we can see that $\Delta M^{H_3}_K$ is still far below the required limit in the taken ranges of $h^{L,R}_{21}$; however, the allowed parameter spaces of $h^{L,R}_{21}$ could be further limited by the required limit of $|\epsilon^{\rm NP}_{K}| \leq 0.4 \times 10^{-3}$. 

 It can be seen from the Fig.~\ref{fig:Box_H3_ab}(b) that when $|h^R_{21}|$ is becoming smaller, the allowed   $|h^L_{21}|$ is becoming  larger due to $C^{LR}_{1,2} \propto h^L_{21} h^R_{21}$. If we take $h^R_{21} \approx 0$, i.e., ${\cal H}^{WH_3}_{\Delta S=2} \approx 0$ and $C^{LR}_{1,2}\approx 0$,  the $h^L_{21}$, dictated by the $Q^{VLL(VRR)}_1$ and $Q^{SLL(SRR)}_{1,2}$ effects, can be much larger than ${\cal O}(10)$. Since $h^{L}_{21}$ is defined through $1/|g^2 V^*_{ts} V_{td}| \sim  6.4 \times 10^{3}$, $h^L_{21}$ of ${\cal O}(30)$ indicates $|g^L_{31}|\sim |g^L_{32}|\sim 0.07$ and is still in the perturbation range. 

  %%%%

\section{ Numerical analysis on $\epsilon'/\epsilon$ in the diquark model}

We numerically study the ${\bf H}_3$ contributions to $\epsilon'/\epsilon$ in this section. Based on the earlier discussions, it is known that three possible mechanisms can contribute to the Kaon direct CP violation, including the tree-level diagram, the QCD and EW penguins, and the chromomagnetic dipole; in addition,  their formulations are given in Eq.~(\ref{eq:ep_eK_T}), Eq.~(\ref{eq:ep_eK_P}), and Eq.~(\ref{eq:epsilon_p1}), respectively. In the following, we discuss their contributions one by one.

\subsection{ Tree-level }

 From $(\epsilon'/\epsilon)_{T}^{H_3}$ shown in Eq.~(\ref{eq:ep_eK_T}),  five free parameters are involved at the tree-level-induced $\Delta S=1$ processes, which are $\zeta^{LL,RR}_{21}$, $\zeta^{RL,LR}_{21}$, and $m_{H_3}$. However, it can be seen that the parameter dependence shows up in the form of $\zeta^{RR}_{21} -\zeta^{LL}_{21}$ and $\zeta^{RL}_{21}-\zeta^{LR}_{21}$; thus, it is more convenient to show the numerical analysis if we use these two forms of parameters as the relevant parameters. In addition, since $\zeta^{\chi}_{21}$ is scaled  by $V^*_{ts} V_{td}$, like the case in $h^{L(R)}_{21}$, where the KM phase is taken as the unique origin of CP violation, we also assume $\zeta^{\chi}_{21}$ to be real parameters in this study although this assumption generally is not necessary. 

\begin{figure}[phtb]
\includegraphics[scale=0.60]{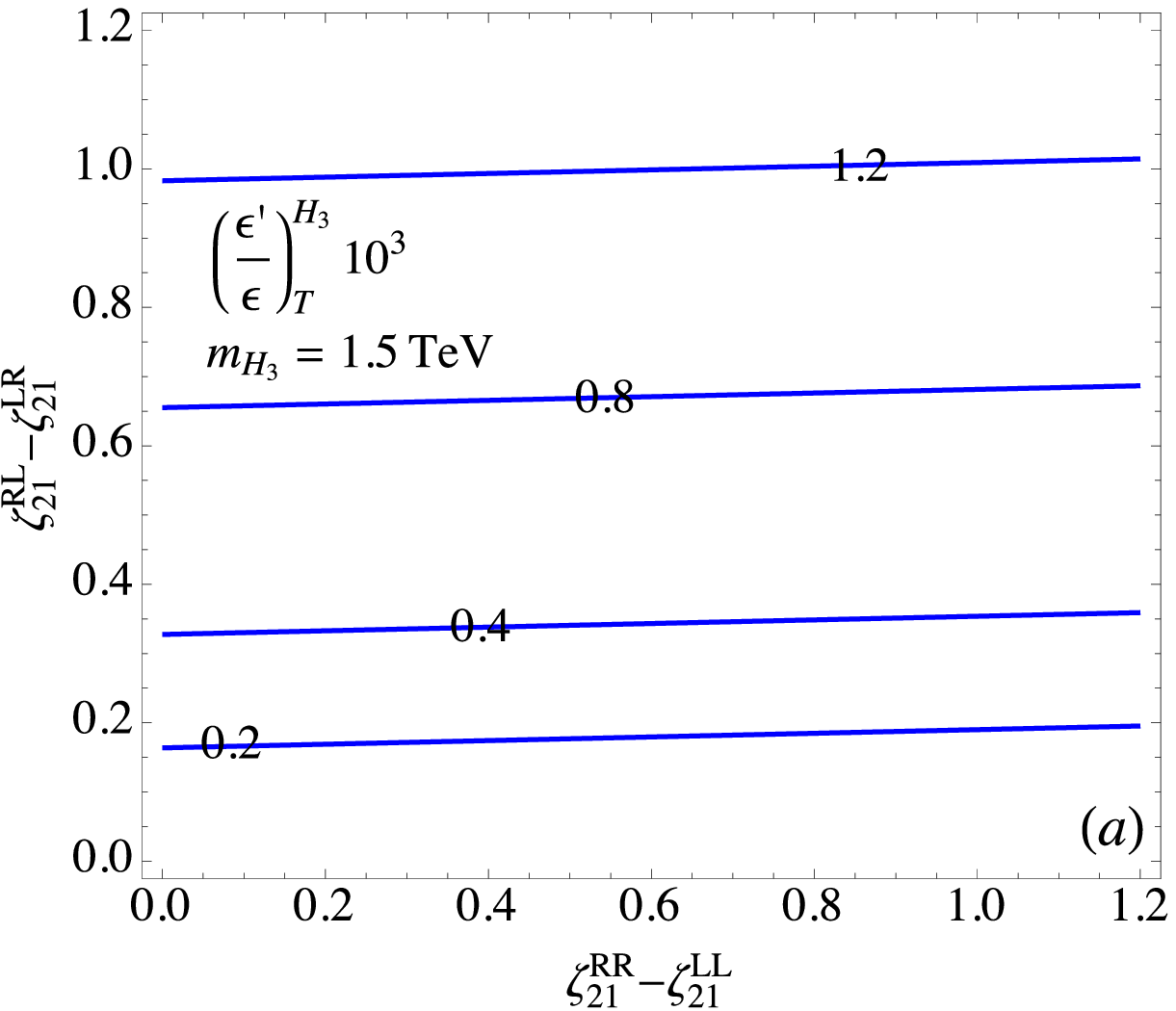}
\includegraphics[scale=0.60]{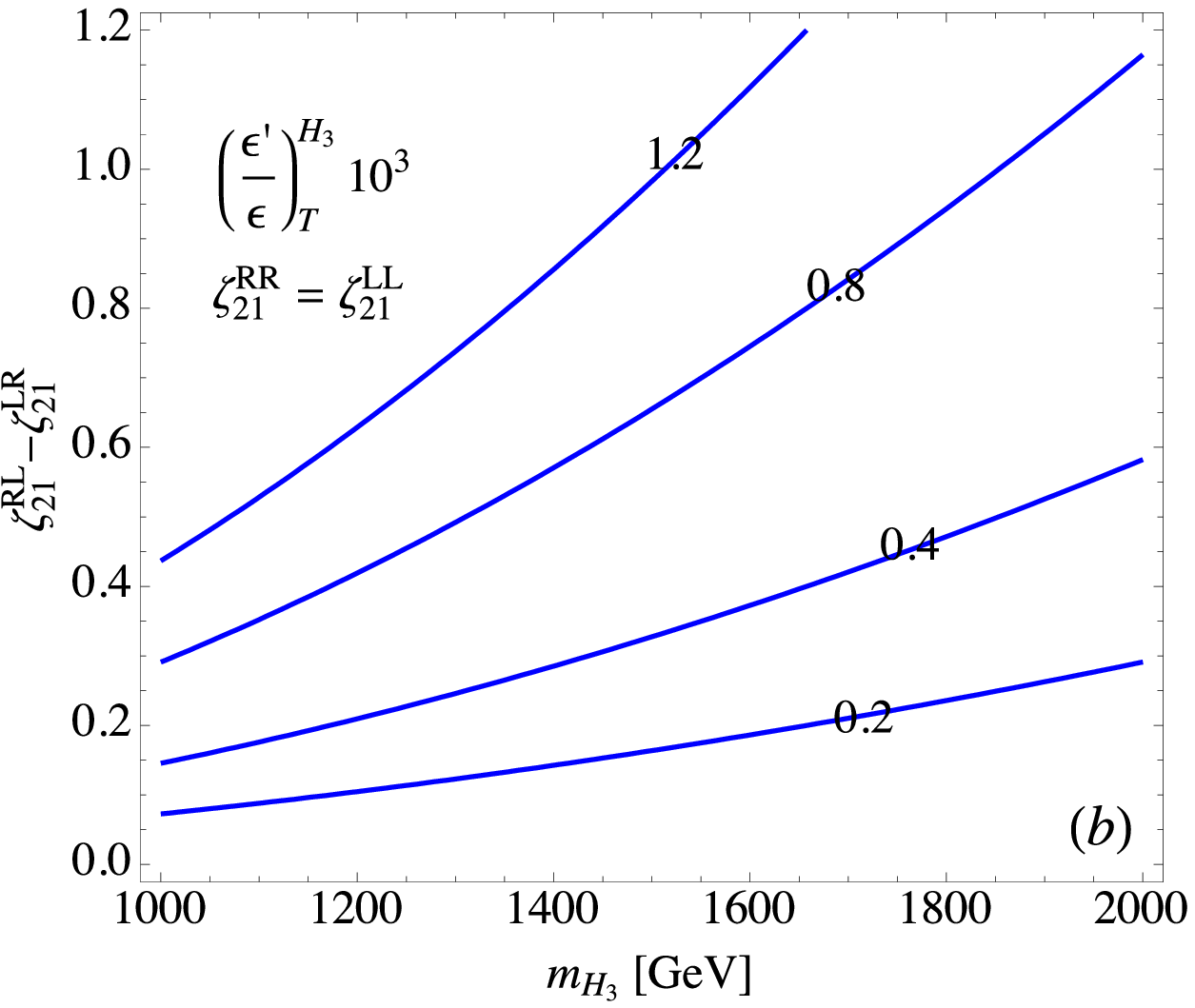}
 \caption{  Contours for $\left( \epsilon'/\epsilon \right)^T_{H_3}$ (in units of $10^{-3}$) as a function of (a) $\zeta^{RR}_{21}-\zeta^{LL}_{21}$ and $\zeta^{RL}_{21} - \zeta^{LR}_{21}$ and (b) $m_{H_3}$ and $\zeta^{RL}_{21} - \zeta^{LR}_{21}$, where $m_{H_3}=1.5$ TeV is used in plot (a), and we assume $\zeta^{RR}_{21}=\zeta^{LL}_{21}$ in plot (b). }
\label{fig:ep_ek_Tree_ab}
\end{figure}

To illustrate the diquark effects, we show the contours for $Re(\epsilon'/\epsilon)^{H_3}_T$ (in units of $10^{-3}$) as a function of  $\zeta^{RR}_{21} -\zeta^{LL}_{21}$ and $\zeta^{RL}_{21}-\zeta^{LR}_{21}$ in Fig.~\ref{fig:ep_ek_Tree_ab}(a), where $m_{H_3}=1.5$ TeV is used. From the plot, $(\epsilon'/\epsilon)^{H_3}_T$ is insensitive to $\zeta^{RR}_{21}-\zeta^{LL}_{21}$. This behavior can be understood from the small coefficient of $ 2 r_1 y_W/z_{-}$ in $T^{1/2}_{H_3}$, where it is above one order of magnitude smaller than $0.67 r_2 y_W/(2 Re A_2)$ in $T^{3/2}_{H_3}$; that is, $T^{3/2}_{H_3}$ dominates the contribution to $(\epsilon'/\epsilon)^{H_3}_T$. Assuming  $\zeta^{RR}_{21}=\zeta^{LL}_{21}$, we show the contours for $(\epsilon'/\epsilon)^{H_3}_T$ as a function of $\zeta^{RL}_{21}-\zeta^{LR}_{21}$ and $m_{H_3}$ in Fig.~\ref{fig:ep_ek_Tree_ab}(b). From these plots, it can be seen that the tree-level diquark effect can significantly enhance $\epsilon'/\epsilon$. 

To further understand the typical size of the $g^{\chi}_{11(12)}$ parameter, we can take $g^{R}_{11}\sim g^{L}_{12}$ and $|\zeta^{RL}_{21}| \sim 0.5$ as an example. Following $\zeta^{RL}_{21} = g^R_{11} g^{L*}_{12}/(g^2 V^*_{ts} V_{td})$, we then obtain $|g^{R}_{11}|\sim |g^{L}_{12}|\sim 0.0088$, which is much smaller than 0.07 the typical value of $g^{\chi}_{31(32)}$ bounded by the $\epsilon^{\rm NP}_K$.

  %%%%

\subsection{QCD and EW penguins}

According to the formulation of $Re(\epsilon'/\epsilon)^{H_3}_P$ in Eqs.~(\ref{eq:ep_eK_P}) and (\ref{eq:a_P}) and the relevant effective Wilson coefficients at $\mu=m_c$ defined in Eq.~(\ref{eq:Delta_yH3}),  the diquark contributions are dictated by the factors $\delta y^{H_3}_{a}$ ($a=3,4,6,7,9$), which exhibit the left-right asymmetry at the $\mu=m_{H_3}$ scale. In order to observe the magnitude of each $\delta y^{H_3}_a$,  following  Eq.~(\ref{eq:WC_Q3-6}) and Eq.~(\ref{eq:WCH3}), we show the $h^{L(R)}_{21}$ dependence with $m_{H_3}=1.5$ TeV as:
 \begin{align}
\delta y^{H_3}_3 & \approx \left(0.81 h^L_{21}+0.04 h^R_{21}\right)\times 10^{-4}\,, \nonumber \\
\delta y^{H_3}_{4}& \approx  0.11 \left( h^L_{21}- h^R_{21}\right)\times 10^{-4}\,, \nonumber \\
\delta y^{H_3}_{5} & \approx \left( -0.04 h^L_{21} + 0.89 h^R_{21} \right)\times 10^{-4}\,, \nonumber \\
\delta y^{H_3}_6 & \approx  0.11 \left(  h^L_{21}- h^R_{21} \right) \times 10^{-4}\,, \nonumber \\
\delta y^{H_3}_{7} & \approx \left(0.71 h^L_{21} - 2.54 h^R_{21} \right) \times 10^{-4}\,, \nonumber\\
\delta y^{H_3}_{9} & \approx \left( -2.69 h^L_{21} + 0.85 h^R_{21} \right) \times 10^{-4} \,. \label{eq:delta_yH3_a}
\end{align}
 Based on the results, we can understand  each $\delta y^{H_3}_a$ as follows: for $\delta y^{H_3}_3$, since there is a $y_W$ suppression factor in the QCD-penguin, the main contribution is from the $Z$-penguin, i.e. $C^Z_3 \propto I_{Z} h^L_{21}$; therefore, it can be seen that the  $h^L_{21}$ part is much larger than the $h^R_{21}$ part. Because  $\delta y^{H_3}_{4(6)}$ is only from the QCD-penguin, it can be seen that $h^{L}_{21}$ and $h^{R}_{21}$ have equal contributions; in addition, since $y^{(\prime)H_3}_{4(6)}$ is a factor of 3 larger than the QCD-penguin part of $y^{(\prime)H_3}_{3}$,  we therefore see that  the $0.11$ factor in $\delta y^{H_3}_{4(6)}$ is almost  a factor of 3 larger than the $0.04$ appearing in the parentheses of $\delta y^{H_3}_{3}$. The behavior of $\delta y^{H_3}_5$ should be similar to $\delta y^{H_3}_3$, but it is dominated by $C'_5 \propto I_{Z} h^R_{21}$.
 %, and due to $I_{ZR}/I_{ZL} \sim -0.2$, 
 %we thus see a smaller value with the  opposite sign in the $h^R_{21}$ part of $\delta y^{H_3}_5$. 
 
 Although $\gamma$- and $Z$-penguin both contribute to  $\delta y^{H_3}_7$, due to the $y_W$ suppression appearing in $\gamma$-penguin, $\delta y^{H_3}_7$ indeed is dominated by the $Z$-penguin. It can be found that the $h^L_{21}$ and $h^R_{21}$ terms in $\delta y^{H_3}_7$ are different from the $h^L_{21}$ term in $\delta y^{H_3}_{3}$ and the $h^R_{21}$ term in $\delta y^{H_3}_5$ by factors of  $4 \sin^2\theta_W\approx 0.92$ and $-4$, respectively. According to these differences, we can roughly understand the numbers in $\delta y^{H_3}_7$ from the corresponding numbers in $\delta y^{H_3}_3$ and $\delta y^{H_3}_5$. From Eq.~(\ref{eq:WC_Q3-6}), $\delta y^{H_3}_9$ is also dominated by the $Z$-penguin. We find that  the $h^L_{21}$ and $h^R_{21}$ terms in $\delta y^{H_3}_{9}$ approximately differ from the corresponding terms in $\delta y^{H_3}_3$ and $\delta y^{H_3}_5$ by factors of $-4 + 4 \sin^2\theta_W \approx -3.08$ and $4\sin^2\theta_W$, respectively.  Using  these factors, we then can roughly obtain the numbers  in the $\delta y^{H_3}_9$ from  those numbers in $\delta y^{H_3}_{3}$ and $\delta y^{H_3}_5$.  
%\begin{align}
%  \Delta y^{H_3}_4 (m_c) & \approx  \left( 3.95 h^L_{21} - 0.06 h^R_{21} \right)\times 10^{-4} \,, \nonumber \\
  %
 %   \Delta y^{H_3}_6 (m_c) & \approx  \left( 2.29 h^L_{21} +0.69 h^R_{21} \right)\times 10^{-4} \,, \nonumber \\
    %
%    \Delta y^{H_3}_8(m_c) & \approx  \left( -4.05 h^L_{21} -2.72  h^R_{21} \right)\times 10^{-4} \,, \nonumber \\
    %
%    \Delta y^{H_3}_9(m_c) & \approx  \left( 16.78 h^L_{21} +1.16 h^R_{21} \right)\times 10^{-4}  \,, \nonumber \\
    %
%     \Delta y^{H_3}_{10}(m_c) & \approx \left( -8.08 h^L_{21} -0.56 h^R_{21} \right)\times 10^{-4} \,,
%  \end{align}

\begin{figure}[phtb]
\includegraphics[scale=0.60]{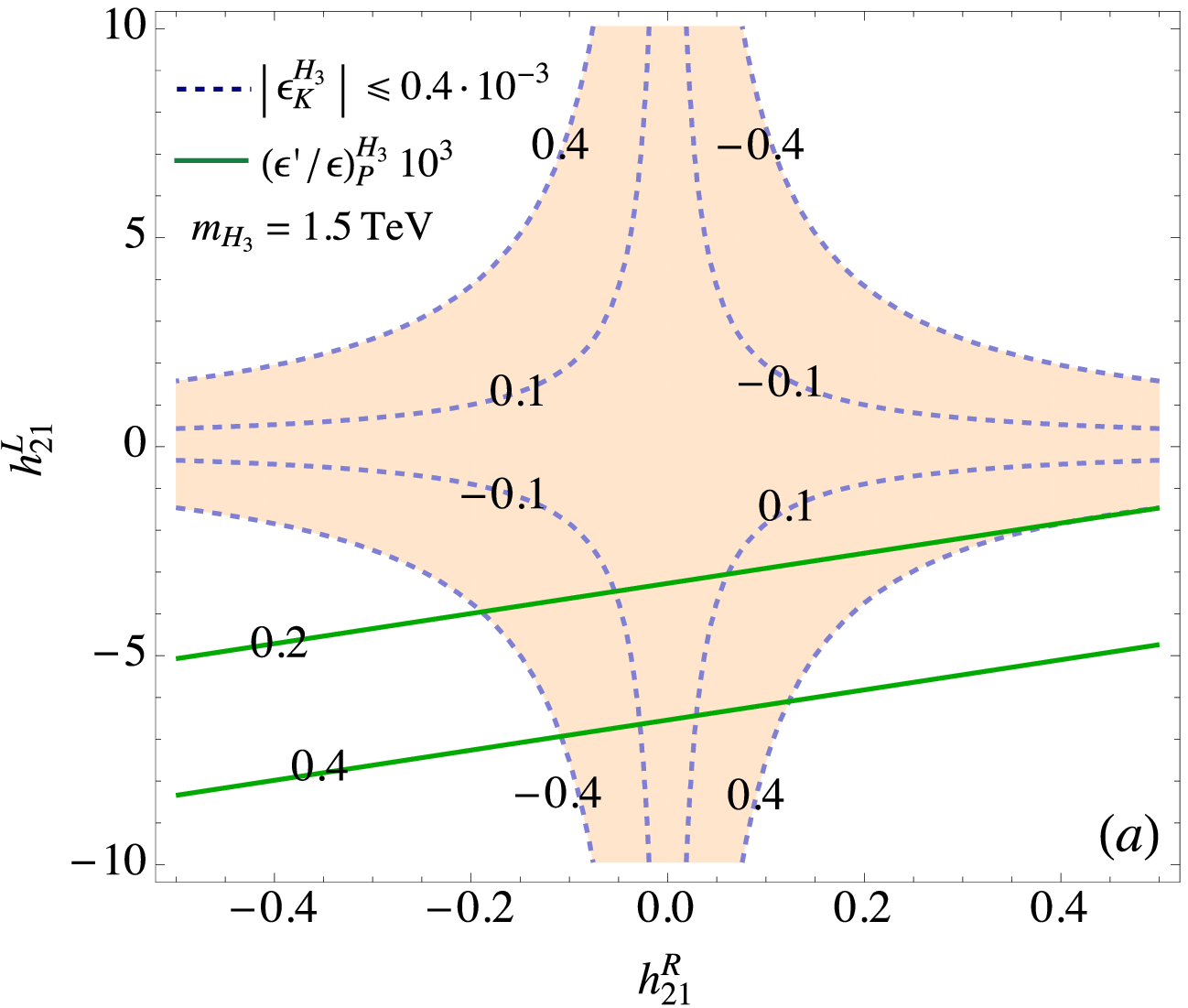}
\includegraphics[scale=0.60]{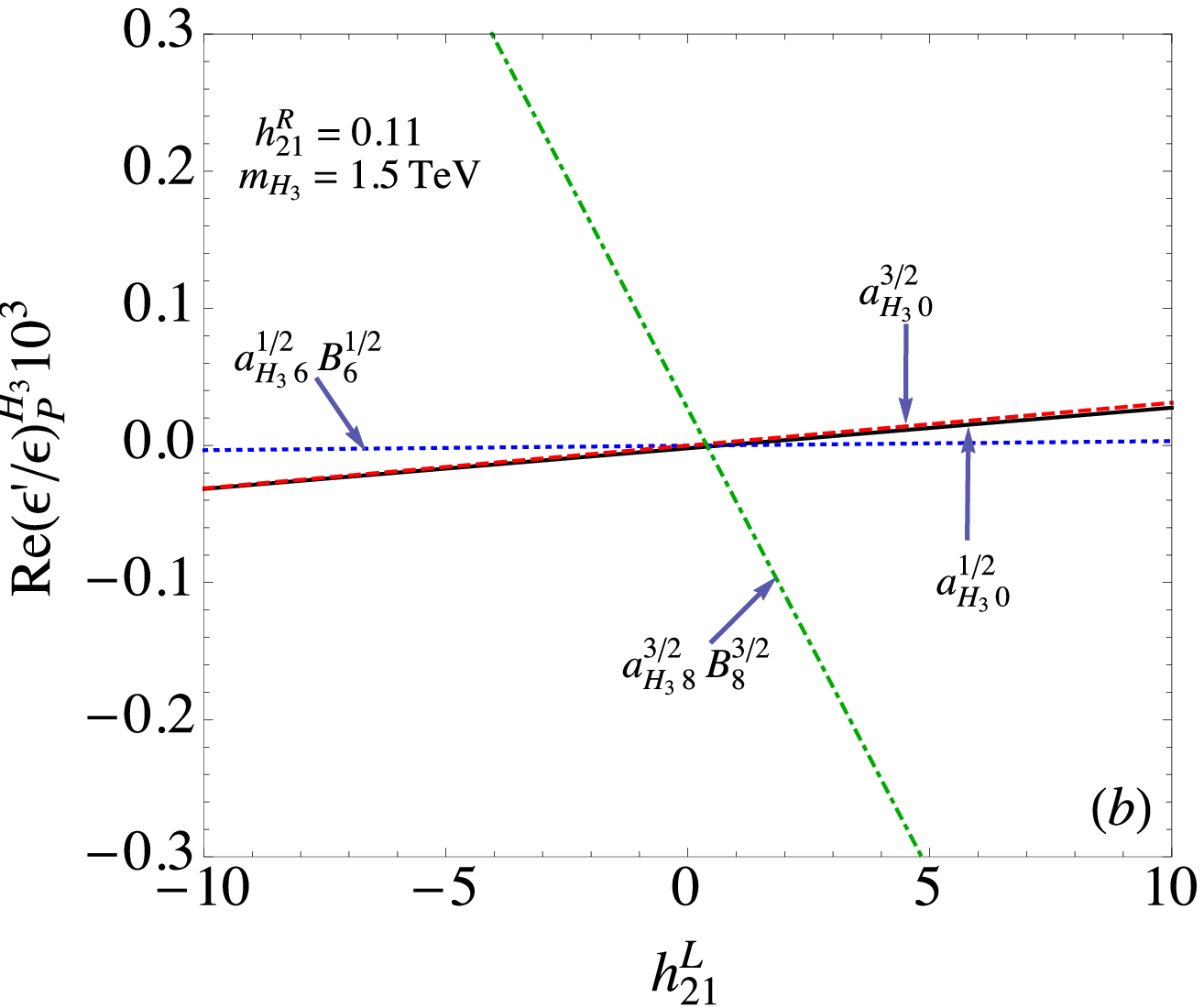}
 \caption{ (a) Contours for $(\epsilon'/\epsilon)^{H_3}_P$ (in units of $10^{-3}$) as a function of $h^L_{21}$ and $h^R_{21}$, where $m_{H_3}=1.5$ TeV is used, and the dashed lines and shaded area denote the constraint from $|\epsilon^{H_3}_K|\leq 0.4\times 10^{-3}$. (b) Each contribution of $a^{1/2}_{H_3 0}$, $a^{1/2}_{H_3 6} B^{(1/2)}_6$, $a^{3/2}_{H_3 0}$, and $a^{3/2}_{H_3 8} B^{3/2}_8$ with $m_{H_3}=1.5$ TeV and $h^R_{21}=0.11$. }
\label{fig:ep_eK_pen_ab}
\end{figure}

Since $m_{H_3}$ is a global parameter in the study, we can simplify the numerical analysis by fixing its value. Hereafter, we  fix $m_{H_3}=1.5$ GeV in the numerical calculations, unless stated otherwise. Thus, we can implement the results in Eq.~(\ref{eq:delta_yH3_a}) to $\Delta y^{H_3}_i(m_c)$ (i=4,6,8,9,10) in Eq.~(\ref{eq:Delta_yH3}).   Using Eqs.~(\ref{eq:ep_eK_P}) and (\ref{eq:a_P}), we plot the contours for $(\epsilon'/\epsilon)^{H_3}_P$ (in units of $10^{-3}$) as a function of $h^L_{21}$ and $h^R_{21}$ in Fig.~\ref{fig:ep_eK_pen_ab}(a), where the shaded area denotes the constraint of  $|\epsilon^{H_3}_K|\leq 0.4 \times 10^{-3}$. From the plot, it can be clearly seen that the diquark parameter spaces, allowed to enhance $\epsilon'/\epsilon$, are still wide when the strict bound from $\epsilon_K$ is included. In order to understand the  role of $a^{1/2}_{H_3 (0, 6)}$ and $a^{3/2}_{H_3 (0, 8)}$, which are defined in Eq.~(\ref{eq:a_P}), in $\epsilon'/\epsilon$, we show each $a^{1/2,3/2}_{H_3 (0,6,8)}$ effect on $Re(\epsilon'/\epsilon)^{H_3}_P$ in Fig.~(\ref{fig:ep_eK_pen_ab})(b), where the solid, dotted, dashed, and dot-dashed lines denote the contributions of $a^{1/2}_{H_3 0}$, $a^{1/2}_{H_3 6}$, $a^{3/2}_{H_3 0}$, and $a^{3/2}_{H_3 8}$, respectively, and $h^R_{21}=0.11$ is taken. Clearly, $a^{3/2}_{H_3 8}$ makes  the main contribution, and  this is because the factor $r_2 \langle Q_8\rangle_2/ReA_2$ in $a^{3/2}_{H_3 8}$ is larger than the others by more than one order of magnitude. In addition, it can be seen that in order to obtain positive $(\epsilon'/\epsilon)^{H_3}_P$,   $h^L_{21}$ prefers  negative values. We can simply understand the preference as follows: It is known that $(\epsilon'/\epsilon)^{H_3}_P$ is dominated by $- a^{2/3}_{H_3 8} \propto - \Delta y^{H_3}_8(m_c) \sim -\delta y^{H_3}_7 \propto  ( - 0.71 h^L_{21} +  2.54 h^R_{21})$. Therefore,   a negative $h^L_{21}$ can positively enhance $(\epsilon'/\epsilon)^{H_3}_P$. 
  %%%%

\subsection{Chromomagnetic dipole}

\begin{figure}[phtb]
\includegraphics[scale=0.60]{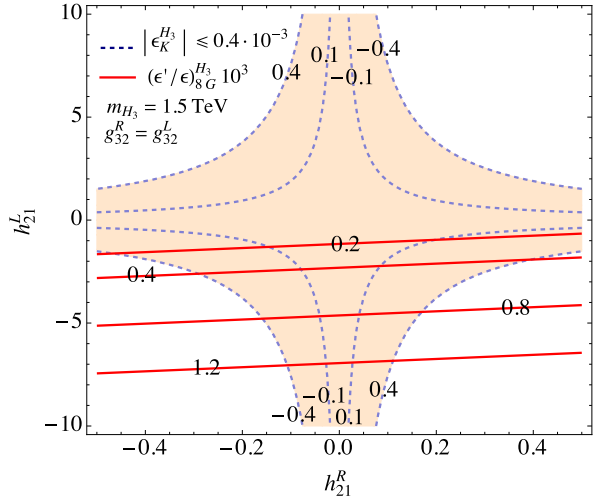}
 \caption{ The legend is the same as that in Fig.~\ref{fig:ep_eK_pen_ab}(a) with the exception of  $(\epsilon'/\epsilon)^{H_3}_{8G}$. }
\label{fig:ep_ek_Q8G}
\end{figure}

From Eq.~(\ref{eq:Cm_8G}), it can be seen that the involved new parameters contributing to $\epsilon'/\epsilon$ through the CMOs are $h^{L,R}_{12}$ and simply appear in the form of $h^R_{21}-h^L_{21}$. With $m_{H_3}=1.5$ TeV, we show the contours for $(\epsilon'/\epsilon)^{H_3}_{8G}$ (in units of $10^{-3}$)
as a function of $h^L_{21}$ and $h^R_{21}$ in Fig.~\ref{fig:ep_ek_Q8G}, where the shaded area denotes the constraint of $\epsilon^{H_3}_{K}\leq 0.4 \times 10^{-3}$. From the results, we can see that the $\epsilon'/\epsilon$  can be significantly enhanced by the CMOs in the diquark model when the bound from the $\epsilon_K$ is satisfied.  Due to the dependence of $h^R_{21}-h^L_{21}$, a negative $h^L_{21}$ can lead to a positive $(\epsilon'/\epsilon)^{H_3}_{8G}$.
Comparing the results with those in $(\epsilon'/\epsilon)^{H_3}_P$,
 it can be found that  $(\epsilon'/\epsilon)^{H_3}_{8G}$ is larger than  $(\epsilon'/\epsilon)^{H_3}_P$ in the same allowed parameter space of $h^L_{21}$. 

  %%%%

\section{Summary}

We investigated the color-triplet diquark ${\bf H}_3$ contributions to the $\Delta S=2$ and $\Delta S=1$ processes in detail. In addition to the ${\bf H}_{3}$ Yukawa couplings to the SM quarks, we also derive the strong and electroweak gauge couplings to ${\bf H}_{3}$. Using the obtained couplings, we calculated renormalized vertex functions for $d \to s (g^{(*)}, \gamma^{(*)}, Z)$.  Based on the results, we studied the implications on the Kaon direct and indirect CP violation. 

We found that the box diagrams mediated by one $W(G)$-boson  and one ${\bf H}_{3}$ for $\Delta S=2$, which were neglected in~\cite{Barr:1989fi}, play an important role on the constraint of the parameter $h^R_{21}$ when the sizable top-quark mass is taken. The constraint on $h^L_{21}$ can be achieved through the purely  ${\bf H}_{3}$-mediated box diagrams.  

It was found that three potential mechanisms  could enhance the Kaon direct CP violation parameter $\epsilon'/\epsilon$, such as the tree-level diagram, the QCD and electroweak penguins, and the chromomagnetic dipole operators. To clearly see each effect, we separately discuss their contributions.  In order to study the $\epsilon'/\epsilon$, in this work, we simply assume that the CP violating origin only arises from the so-called KM phase of the CKM matrix in the SM. Using the limited parameters and the hadronic matrix elements provided in~\cite{Aebischer:2018rrz}, we find that the $\Delta S=2$ process cannot give a strict bound on the tree-level parameters $\zeta^{RR,LL}_{21}$ and $\zeta^{RL,LR}_{21}$; therefore, the  parameter spaces  to significantly  enhance $(\epsilon'/\epsilon)$ are wide. 

The parameters associated with the QCD and electroweak penguins and the chromomagnetic dipole are the same.
Although these parameters used to enhance $\epsilon'/\epsilon$ are bounded by the Kaon  indirect CP violation $\epsilon_K$, it was found that $\epsilon'/\epsilon$ can still be significantly enhanced by these mechanisms. In addition, in the same  parameter space of $h^L_{21}$, which can generate a sizable $\epsilon'/\epsilon$, the contribution to $\epsilon'/\epsilon$ from the chromomagnetic operators is larger than that from the QCD and EW penguins.

\appendix

\section{}

\subsection{Renormalized two- and three-point diagrams for gluon emission}

To deal with the calculations of one-loop Feynman diagrams, we show the useful 
$d$-dimensional integral  as:
 \begin{align}
J(d, m,n,\mu^2_B) & = \int \frac{d^d \ell }{(2\pi)^d} \frac{(\ell^2)^m}{(\ell^2 - \mu^2_{B})^n }  \nonumber \\
& =i \frac{(-1)^{m-n} (\mu^2_B)^{d/2+m-n}}{(4\pi)^{d/2}} \frac{\Gamma(n-m-d/2)\Gamma(m+d/2)}{\Gamma(d/2) \Gamma(n)}\,.
 \end{align}
Using dimensional regulation with $d=4+2\epsilon$, renormalization scale $\mu$, and $\Gamma(-\epsilon)=-1/\epsilon - \gamma_E$, the relevant integrals in the study are explicitly written as:
 \begin{align}
 J(d,0,2,\mu^2_{B}) &  = i \frac{\mu^{2\epsilon} }{(4\pi)^2 }  \ln\frac{\Lambda^2}{\mu^2_{B}}\,, \nonumber \\
  J(d,0,3,\mu^2_{B}) & = -i \frac{1}{(4\pi)^2 \Gamma(3) } \frac{1}{\mu^2_{B}} \,, \nonumber \\
  J(d,1,3,\mu^2_{B}) & =  \frac{d}{4} J(d,0,2,\mu^2_{B}) \,, \label{eq:ints}
 \end{align}
where we define $\ln\Lambda^2 = -1/\epsilon -\gamma_{E} + \ln (4\pi\mu^2)$, and $\gamma_{E}$ is the Euler-Mascheroni constant. 

   %%%%
\begin{figure}[phtb]
\includegraphics[scale=0.75]{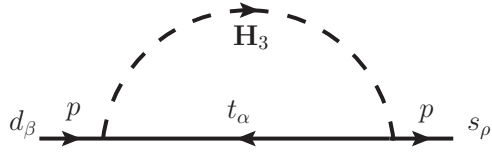}
 \caption{ Self-energy diagram for the $d\to s$ transition mediated by color-triplet diquark ${\bf H}_{3}$. }
\label{fig:SE}
\end{figure}

The  self-energy diagram mediated by ${\bf H}_3$ for the $d\to s$ transition is sketched in Fig.~\ref{fig:SE}. Using the Yukawa couplings in Eq.~(\ref{eq:Yukawa}), the result of Fig.~\ref{fig:SE} can be expressed as:
\begin{align}
i \Sigma(p) &= \bar s \, \Gamma \, d = \bar s \left[ \slashed{p} \chi^V_{21}  
\int^1_0 dx\,  x J(d,0,2,\mu^2_{B1}(p^2) ) \right. \nonumber \\
& \left.  + m_t  \chi^S_{21} \int^1_0 dx\,   J(d,0,2,\mu^2_{B1}(p^2) ) \right] d  \,, \label{eq:Sigma_p}\\
\chi^V_{21} & =  g^{L*}_{32} g^L_{31} P_L + g^{R*}_{32} g^{R}_{31} P_R \,, \nonumber \\
\chi^S_{21} & = g^{R*}_{32} g^L_{31} P_L +  g^{L*}_{32} g^R_{31} P_R \,,  \label{eq:chiVS}
\end{align}
where $( K^a)^{\rho\alpha} (\bar K_a)_{\alpha\beta} = \delta_\beta^{\rho}$ is used, and $\mu^2_{B1}(p^2) = m^2_{H_3} x + m_t^2 (1-x) - p^2 x (1-x) $. To obtain the renormalized $\Gamma$, we require $\Sigma(p)=0$ when the momentum of the external quark is taken on the mass shell, i.e., $p=p_d$ or $p=p_s$. If we write the renormalized $\Gamma_R$ as:
 \begin{equation}
 \Gamma_R = \Gamma + C_1 \slashed{p} P_L + C_2 \slashed{p} P_R + C_3  P_R + C_4   P_L\,, \label{eq:G_R}
 \end{equation}
 the requirements of $\Sigma_R(p_d)=0$ and $\Sigma_R(p_s)=0$ lead to 
  \begin{align}
  C_1 \simeq - g^{L*}_{32} g^L_{31}  \int^1_0 dx\, x J(d,0,2,\mu^2_{B1}(0))\,, \nonumber \\
  C_2 \simeq  - g^{R*}_{32} g^R_{31}  \int^1_0 dx\, x J(d,0,2,\mu^2_{B1}(0))\,, \nonumber \\
C_3 \simeq - g^{L*}_{32} g^R_{31} m_t  \int^1_0 dx\,  J(d,0,2,\mu^2_{B1}(0))\,, \nonumber \\
 C_4 \simeq  - g^{R*}_{32} g^L_{31}  m_t \int^1_0 dx\,  J(d,0,2,\mu^2_{B1}(0))\,,
  \end{align}
  where we have dropped the light quark mass effects. We note that the mass dimension in $C_{1(2)}$ is  different from that in $C_{3(4)}$.

The color-triplet-mediated three-point diagrams for $d\to s g^{(*)}$ are shown in Fig.~\ref{fig:penguin}, where $g^{(*)}$ denotes the on-shell (off-shell) gluon. The result of  Fig.~\ref{fig:penguin}(a), where the gluon is emitted from the top-quark, is given as:
 \begin{align}
 i \Gamma^{A\mu}_{a} & = g_s (K^b)^{\rho \sigma} (T^A)^{\alpha}_{\sigma} (\bar K_{b})_{\alpha \beta} \Gamma(3) \int^1_0 dx_1 \int^{x_1}_{0} dx_2 \int  \frac{d^d \ell}{(2\pi)^d} \frac{1}{(\ell^2 - \mu^2_{B2}(k^2) )^3} \nonumber \\
 & \times \bar s_\rho \left\{ - A^\mu_1 \chi^V_{21}  +   A^\mu_2 \chi^S_{21} \right\} d^\beta \,, \label{eq:Gamma_Aa}\\
 A^\mu _1 & = \slashed{\ell} \gamma^\mu \slashed{\ell}+[m^2_t -  k^2 x_2 (x_2-x_1) ] \gamma^\mu  \,, \nonumber \\
 A^\mu_2 & = m_t [(x_2 -x_1)  \gamma^\mu \slashed{k} + x_2  \slashed{k} \gamma^\mu] \,, \nonumber \\
 \mu^2_{B2}(k^2) & = m^2_{H_3} (1-x_1) + m^2_t x_1 + k^2 x_2 (x_2 -x_1)\,, \label{eq:A1A2}
 \end{align}
 where $T^A$ are the generators of $SU(3)_C$ and their normalizations are taken as $Tr(T^A T^B) = \delta^{AB}/2$. We find that the color factor and the ultraviolet divergent  part can be  expressed as: 
 \begin{align}
  (K^b)^{\rho \sigma} (T^A)^{\alpha}_{\sigma} (\bar K_{b})_{\alpha \beta}  & = -\frac{(T^A)^{\rho}_{\beta}}{2} \,, \nonumber \\
 \int \frac{d^d \ell }{(2\pi)^d} \frac{\slashed{\ell} \gamma^\mu \slashed{\ell}}{(\ell^2 - \mu^2_{B_2}(k^2))^3} & = -\gamma^\mu \frac{1+\epsilon }{\Gamma(3)}  J(d,0,2,\mu^2_{B_2}(k^2))\,.
  \end{align}
Accordingly, $\Gamma^{A\mu}_a$ can be reformulated as:
 \begin{align}
 i \Gamma^{A\mu}_{a} & =- i \frac{g_s}{2 (4\pi)^2}   \bar s  \gamma^\mu \chi^V_{21} T^A d \int^1_0 dx_1 \int^{x_1}_{0} dx_2   \nonumber \\
 & \times  \left[ \mu^{2\epsilon} \left( \ln\frac{\Lambda^2}{\mu^2_{B2}(k^2)} -1\right)  + \frac{m^2_t + k^2 x_2 (x_2 -x_1)}{\mu^2_{B2}(k^2)}\right]  \nonumber  \\
 &  + i \frac{g_s }{2 (4\pi)^2}   \bar s A^\mu_2 \chi^S_{21}  T^A d \int^1_0 dx_1 \int^{x_1}_{0} dx_2  \frac{1}{\mu^2_{B2}(k^2)}\,. \label{eq:penguin_a}
 \end{align}

Using the diquark-gluon coupling shown in Eq.~(\ref{eq:gH3H3}), the result of Fig.~\ref{fig:penguin}(b), where the gluon is emitted from the ${\bf H_3}$, can be obtained as:
 \begin{align}
 i\Gamma^{A\mu}_{b} &= g_s (K^a)^{\rho \alpha}  (\bar K_{b})_{\alpha \beta}  (t^A)^b_a \Gamma(3) \int^1_0 dx_1 \int^{x_1}_{0} dx_2 \int  \frac{d^d \ell}{(2\pi)^d} \frac{1}{(\ell^2 - \mu^2_{B3}(k^2) )^3} \nonumber \\
 & \times \bar s_\rho \left\{  - B^\mu_1 \chi^V_{21}  +   B^\mu_2  \chi^S_{21}  \right\} d^\beta \,, \label{eq:Gamma_Ab} \\
 B^\mu_1 & = 2 \ell^\mu \slashed{\ell} \,,~~
 B^\mu_2 = m_t [p^\mu_s (2-2x_1) - k^\mu (1-2 x_1 + 2 x_2)] \,, \nonumber \\
 \mu^2_{B3}(k^2) &= m^2_{H_3} x_1 + m^2_t (1-x_1) + k^2 x_2 (x_2 - x_1)\,, \label{eq:B1B2}
 \end{align}
 where $(t^A)^{b}_{a} = 2 Tr (\bar K_a T^A K^b)$ denotes the effective color factor.
 Similar to Eq.~(\ref{eq:Gamma_Aa}), the color factor and ultraviolet divergent part of Fig.~\ref{fig:penguin}(b) can be obtained as:
  \begin{align}
  (K^a)^{\rho \alpha} (\bar K_b)_{\alpha \beta} (t^A)^b_a & = 2 (K^a)^{\rho \alpha} (\bar K_b)_{\alpha \beta} Tr \bar K_a T^A K^b = \frac{(T^A)_\beta^{\rho}}{2}\,, \nonumber \\
   \int \frac{d^d \ell }{(2\pi)^d} \frac{\ell^\mu \slashed{\ell}}{(\ell^2 - \mu^2_{B_3}(k^2))^3} &= \frac{\gamma^\mu }{2 \Gamma(3)}   J(d,0,2,\mu^2_{B3})\,.
  \end{align}
Thus, the vertex function for the gluon emitting from the diquark is given by:
\begin{align}
 i\Gamma^{A\mu}_{b} &= -i \frac{g_s}{2 (4\pi)^2} \bar s  \gamma^\mu \chi^V_{21}  T^A d \int^1_0 dx_1 \int^{x_1}_{0} dx_2  \mu^{2\epsilon} \ln\frac{\Lambda^2}{\mu^2_{B3}(k^2)} \nonumber \\
 & -  i \frac{g_s}{2 (4\pi)^2} \bar s  \chi^S_{21}  T^A d   \int^1_0 dx_1 \int^{x_1}_{0} dx_2  \frac{B^\mu_2}{\mu^2_{B3}(k^2)} \,.\label{eq:penguin_b}
 \end{align}

 From the Ward-Takahashi identity, it is known that the three-point vertex correction can be related to  the two-point function $\Sigma(p) = \bar s \Gamma d$ through the relation:
  \begin{equation}
  k_\mu \Gamma^{A\mu}=k_\mu \Gamma^{A\mu}_{a+b} = g_s (T^A)_\beta^{\rho } \left[ \Sigma(p-k)_{\rho}^{\beta} -\Sigma(p)_{\rho}^{ \beta}\right],
  \end{equation}
with $\Sigma(p)_{\rho}^{\beta}=\bar s_\rho \Gamma d^\beta$.  In order to obtain the renormalized $\Gamma^{A\mu}$, we can require that the Ward-Takahashi identity is retained as $ k_\mu \Gamma^{A\mu}_{R} = g_s (T^A)^{\rho}_{\beta} \left[ \Sigma_R(p-k)_{\rho}^{\beta} -\Sigma_R(p)_{\rho}^{ \beta}\right]$~\cite{Chia:1983hd,Davies:1991jt}. If we set $\Gamma^{A\mu}_R = \Gamma^{A\mu} + X^{A\mu}$,  the Ward-Takahashi identity can lead to:
 \begin{align}
 X^{A\mu} &=   \bar s \gamma^\mu \chi^V_{21} T^A d \int^1_0 dx \, x J(d,0,2,\mu^2_{B1}(0)) \nonumber \\
 & =  \frac{i}{(4\pi)^2} \bar s \gamma^\mu \chi^V_{21} T^A d \int^1_0 dx \, x \mu^{2\epsilon} \ln\frac{\Lambda^2}{\mu^2_{B1}(0)}\,. \label{eq:XA}
 \end{align}
The ultraviolet divergence of $\Gamma^{A\mu}_R$, which is related to $\ln\Lambda^2$ terms, can then be cancelled as:
 \begin{align}
 \Gamma^{A\mu}_R \Big|_{\rm div} = \Gamma^{A\mu}_{a+b}\Big|_{\rm div} + X^{A\mu}\Big|_{\rm div} \propto  - \int^1_0 dx_1 \int^{x_1}_0 dx_2 \frac{\ln\Lambda^2}{2} \times 2 +  \int^1_0 dx \, x  \ln\Lambda^2 =0\,. 
 \end{align}
In order to verify the gauge invariance, we can take $k^2=0$ for the on-shell gluon; thus,  the Ward identity can be satisfied as:
 \begin{align}
 k_\mu \Gamma^{A\mu}_R & \propto  \int^1_0 dx_1 \int^{x_1}_0 dx_2  \frac{1}{2} \left[  \left (   \ln\frac{\mu^2_{B2}(0)}{m^2_{H_3}} +1 \right) - \frac{m^2_t}{\mu^2_{B2}(0)} + \ln\frac{\mu^2_{B3}(0)}{m^2_{H_3}} \right] \nonumber \\
 &+ \int^1_0 dx \, x\, \ln\frac{\mu^2_{B1}(0)}{m^2_{H_3}} \nonumber \\
 & = \frac{1}{4} + \frac{1}{2} \int^1_0 dx \left[ (1-2x) \ln( x + y_t (1-x)) - \frac{y_t(1-x) }{x+ y_t (1-x)}\right] =0 \,, \label{eq:gaugeQCD}
 \end{align}
with $y_t=m^2_t /m^2_{H_3}$. For $k^2\neq 0$, due to $k^2 \ll m^2_t$, the leading $k^2$ term and chromomagnetic dipole effect of $\Gamma^{A\mu}_R$ can be obtained as:
 \begin{align}
 i \epsilon^{A}_\mu \Gamma^{A\mu}_R  & = - i \frac{g_s k^2}{(4\pi)^2 m^2_{H_3}} I_{G1}(y_t) \bar s \slashed{\epsilon}^A \chi^V_{21} T^A d + i \frac{g_s}{(4\pi)^2} \frac{m_t}{4m^2_{H_3}} I_{G2}(y_t) \bar s \sigma^{\mu \nu} \chi^S_{21} T^A d\, G^A_{\mu\nu}\,, \label{eq:RVg}
 \end{align}
 where the loop-integral functions are given as:
 \begin{align}
 I_{G1}(y) &= \frac{2y^2 + 11 y -7 }{36(1-y)^3} + \frac{(y^3 + 3 y-2) \ln y}{12(1-y)^4} \,, \nonumber \\
 I_{G2}(y) & = - \frac{1}{(1-y)} - \frac{\ln y}{(1-y)^2}\,. \label{eq:loop_fg}
 \end{align}

\subsection{ Renormalized three-point vertex function for $d\to s \gamma^{(*)}$}

In addition to the gluon-penguin diagrams, the electroweak penguin diagrams, i.e. $d\to s \gamma^* (Z^*)$, also make significant contributions to $\epsilon'/\epsilon$. Since photon is a massless particle, like the case in the $d\to s g^*$ process, the leading effect for the  photon emission  $d \to s \gamma^*$ decay should be proportional to $k^2$,  so that the off-shell photon propagator of $1/k^2$  in the $d\to s q\bar q$ processes can be cancelled. 
%Comparing with the photon emission process,  the results of the penguin-induced $d\to s Z^*$  are much easier to obtain.  
 Due to the similarity to the gluon case, in this subsection, we first discuss  the $d\to s \gamma^{(*)}$ process.

The Feynman diagrams for $d \to s \gamma^{(*)}$ are shown in Fig.~\ref{fig:AZ-penguin}. It can be seen that with the exception of gauge couplings, 
the calculations for $d\to s \gamma^{(*)}$ are similar to those for $d\to s g^{(*)}$; therefore, the results of Fig.~\ref{fig:AZ-penguin}(a) and (b) can be respectively obtained from Eqs.~(\ref{eq:penguin_a}) and (\ref{eq:penguin_b}), when the strong interactions are replaced by the electromagnetic interactions. Thus, using $(K^a)^{\alpha \beta} (\bar K_{a})_{\rho \alpha} = \delta^\beta_\rho$ and gauge coupling in Eq.~(\ref{eq:photonH3H3}), the results of Fig.~\ref{fig:AZ-penguin}(a) and (b) can be  formulated as:
\begin{align}
 i \Gamma^{\mu}_{\gamma a} & = i \frac{e_t e }{ (4\pi)^2}   \bar s  \gamma^\mu \chi^V_{21} d \int^1_0 dx_1 \int^{x_1}_{0} dx_2   \nonumber \\
 & \times  \left[ \mu^{2\epsilon} \left( \ln\frac{\Lambda^2}{\mu^2_{B2}(k^2)} -1\right)  + \frac{m^2_t + k^2 x_2 (x_2 -x_1)}{\mu^2_{B2}(k^2)}\right]  \nonumber  \\
 &  - i \frac{e_t e }{ (4\pi)^2}   \bar s A^\mu_2 \chi^S_{21}   d \int^1_0 dx_1 \int^{x_1}_{0} dx_2  \frac{1}{\mu^2_{B2}(k^2)}\,.
 \end{align}
 
 \begin{align}
 i\Gamma^{\mu}_{\gamma b} &= - i \frac{e_t e }{ (4\pi)^2}  \bar s  \gamma^\mu \chi^V_{21}   d \int^1_0 dx_1 \int^{x_1}_{0} dx_2  \mu^{2\epsilon} \ln\frac{\Lambda^2}{\mu^2_{B3}(k^2)} \nonumber \\
 & -  i \frac{e_{H_3} e}{ (4\pi)^2} \bar s   \chi^S_{21}   d   \int^1_0 dx_1 \int^{x_1}_{0} dx_2  \frac{B^\mu_2}{\mu^2_{B3}(k^2)} \,, 
 \end{align}
where $\chi^{V(S)}_{21}$, $A^\mu_2$, and  $B^\mu_2$  can be found in Eqs. (\ref{eq:chiVS}),  (\ref{eq:A1A2}), and (\ref{eq:B1B2}), respectively. In order to obtain the renormalized vertex function, we require that the Ward-Takahashi identity, which is defined as:
\begin{equation}
k_\mu \Gamma^\mu_\gamma = k_\mu (\Gamma^\mu_{\gamma a} + \Gamma^\mu_{\gamma b})= e_d e [\Sigma(p-k) - \Sigma(p)]\,, 
\end{equation}
 is retained when we renormalize the three-point vertex corrections, i.e., $k_\mu \Gamma^\mu_{\gamma R} =  e_d e [\Sigma_R(p-k) - \Sigma_R(p)]$, where $e_d$ denotes the electric charge of a down-type quark, and $\Sigma_R(p)$ can be obtained from Eq.~(\ref{eq:G_R}). If we set $\Gamma^\mu_{\gamma R} = \Gamma^\mu_\gamma + X^\mu_\gamma$, the renormalization requirement can lead to:
  \begin{equation}
  X^\mu_\gamma =  i \frac{e_d e}{(4\pi)^2} \bar s \gamma^\mu \chi^V_{21} d  \int^1_0 dx\, x\, \mu^{2\epsilon} \ln\frac{\Lambda^2}{\mu^2_{B1}(0)}\,. 
  \end{equation}
 Similar to the gluon penguin, the $\ln\Lambda^2$   ultraviolet divergence related terms in $\Gamma^\mu_R$ can be cancelled as:
 \begin{align}
 \Gamma^{\mu}_{\gamma R} \Big|_{\rm div} & = \Gamma^{\mu}_{\gamma(a+b)}\Big|_{\rm div} + X^{\mu}_\gamma \Big|_{\rm div} \nonumber \\
 & \propto  (e_t -e_{H_3})  \int^1_0 dx_1 \int^{x_1}_0 dx_2 \ln\Lambda^2  + e_d  \int^1_0 dx \, x  \ln\Lambda^2 =0\,. 
 \end{align}
 We can verify the $U(1)_{\rm em}$ gauge invariance through the case of $k^2=0$ as:
  \begin{align}
 k_\mu \Gamma^{\mu}_{\gamma R} & \propto  e_t  \int^1_0 dx_1 \int^{x_1}_0 dx_2  \left[  \left ( -  \ln\frac{\mu^2_{B2}(0)}{m^2_{H_3}} -1 \right) + \frac{m^2_t}{\mu^2_{B2}(0)} \right]  \nonumber \\
 &+ e_{H_3} \int^1_0 dx_1 \int^{x_1}_0 dx_2 \ln\frac{\mu^2_{B3}(0)}{m^2_{H_3}}- e_d \int^1_0 dx \, x\, \ln\frac{\mu^2_{B1}(0)}{m^2_{H_3}} \nonumber \\
 & = -\frac{1}{3} - \frac{2}{3} \int^1_0 dx \left[ (1-2x) \ln( x + y_t (1-x)) - \frac{y_t(1-x) }{x+ y_t (1-x)}\right] =0 \,. \label{eq:gaugeQED}
 \end{align}
 Hence, the leading $k^2$ and electromagnetic dipole effects of $\Gamma^{\mu}_R$ can be obtained as:
 \begin{align}
 i \epsilon_\mu \Gamma^{\mu}_{\gamma R}  & = - i \frac{e k^2}{3 (4\pi)^2 m^2_{H_3}} I_{\gamma 1}(y_t) \bar s \slashed{\epsilon} \chi^V_{21}  d - i \frac{e}{(4\pi)^2} \frac{m_t}{6m^2_{H_3}} I_{\gamma 2}(y_t) \bar s \sigma^{\mu \nu} \chi^S_{21}  d\, F_{\mu\nu}\,, \label{eq:L_gamma}
 \end{align}
 where the loop-integral functions are given as:
 \begin{align}
 I_{\gamma 1}(y) &= \frac{25 y^2 -65 y +34 }{36(1-y)^3} + \frac{y^3 + 2( 2-3 y)}{6(1-y)^4}   \ln y\,, \nonumber \\
 I_{\gamma 2}(y) & = -\frac{7-y}{2 (1-y)^2} - \frac{(2+y)\ln y}{(1-y)^3}\,. \label{eq:I_gamma}
 \end{align}

\subsection{ Renormalized three-point vertex function for $d\to s Z^*$}

 To calculate  the $Z$-penguin induced  three-point vertex for $d \to s Z^*$, we write the $Z$-couplings to quarks as:
 \begin{align}
 {\cal L}_{Zqq} & = - \frac{ g}{\cos\theta_W} \bar q \gamma_\mu (C^q_L P_L + C^q_R P_R) Z^\mu \,, \\
 C^q_L & = I^q - e_q \sin^2\theta_W \,, ~ C^q_R = - e_q \sin^2\theta_W\,, \label{eq:CqLR}
 \end{align}
 where $I^q$ and $e_q$ are  the weak isospin  and electric charge of  the $q$-quark, respectively. From the $Z$-boson interactions, it can be seen that the $e_t \sin^2\theta_W$ related currents  indeed are the same as the electromagnetic currents; that is, the corresponding three-point vertex function should be proportional to $k^2$. Since $Z$-boson is a massive particle, unlike the case in $d\to s \gamma^*$, the $k^2$-related effects will be suppressed by $k^2/m^2_Z$ in the decays such as $d\to s q\bar q$ and $d\to s \ell \bar\ell$. Thus, we expect that the renormalized $d \to s Z^*$ vertex is only related to the weak isospin $I^t=1/2$ when the $k^2$ effects are neglected.  In the following, we show the calculated results. 
 
 Using the Yukawa couplings shown in Eq.~(\ref{eq:Yukawa}), the relation $(K^a)^{\rho \alpha} (\bar K_a)_{\alpha \beta}=\delta^\rho_\beta$, and the integrals in Eq.~(\ref{eq:ints}), the results of Fig.~\ref{fig:AZ-penguin}(a) and (b) can be respectively expressed as:
 \begin{align}
 i \Gamma^{\mu}_{Z a} & = i \frac{g  }{ (4\pi)^2 \cos\theta_W }   \int^1_0 dx_1 \int^{x_1}_{0} dx_2   \nonumber \\
 & \times  \left[ \mu^{2\epsilon} \left( \ln\frac{\Lambda^2}{\mu^2_{B2}(0)} -1\right)  \bar s  \gamma^\mu \xi^t_{21} d + \frac{m^2_t }{\mu^2_{B2}(0)} \bar s  \gamma^\mu \eta^t_{21} d \right]  \,,  \\
  i\Gamma^{\mu}_{Z b} &= i \frac{g e_{H_3} \sin^2\theta_W}{ (4\pi)^2 \cos\theta_W } \bar s  \gamma^\mu \chi^V_{21}   d \int^1_0 dx_1 \int^{x_1}_{0} dx_2  \mu^{2\epsilon} \ln\frac{\Lambda^2}{\mu^2_{B3}(0)} \,,  \\
  \xi^q_{21} & = g^L_{31} g^{L*}_{21} C^q_L P_L + g^R_{31} g^{R*}_{32} C^q_R P_R\,, \nonumber \\
  \eta^q_{21} &= g^L_{31} g^{L*}_{21} C^q_R P_L + g^R_{31} g^{R*}_{32} C^q_L P_R\,, \nonumber
 \end{align}
 where we have taken $k^2=0$ and dropped the small effect from the dipole operators. 

The Ward-Takahashi identity for the $d \to s Z^*$ vertex is given by:
 \begin{align}
 k_\mu \Gamma^\mu_Z =k_\mu (\Gamma^\mu_{Za} + \Gamma^\mu_{Zb}) = \frac{g}{\cos\theta_W} \left(\Sigma'(p-k) -\Sigma'(p)\right)\,, \label{eq:WT_Z}
 \end{align}
 where $\Sigma'(p)$ can be obtained from the $\Sigma(p)$ in Eq.~(\ref{eq:Sigma_p}) using $\xi^d_{21}$ instead of $\chi^V_{21}$. If  the renormalized  $\Gamma^\mu_Z$ is written as $\Gamma^\mu_{ZR} = \Gamma^\mu_{Z} + X^\mu_Z$, by requiring $\Gamma^\mu_{ZR}$ to obey the same Ward-Takahashi identity as shown in Eq.~(\ref{eq:WT_Z}),  $X^\mu_Z$ can be found as:
\begin{equation}
X^\mu_{Z} = i \frac{g}{(4\pi)^2 \cos\theta_W} \bar s \gamma^\mu \xi^d_{21} d  \int^1_0 dx\, x\,  \mu^{2\epsilon} \ln\frac{\Lambda^2}{\mu^2_{B1}(0)}
\end{equation}
Thus, we can check the UV divergence-free  as follows:
 \begin{align}
 \Gamma^\mu_{ZR}\Big|_{\rm div} =\Gamma^\mu_{Z(a+b)}\Big|_{\rm div} + X^\mu_Z\Big|_{\rm div} & \propto  \bar s \gamma^\mu P_L d\, \ln\Lambda^2  g^L_{31} g^{L*}_{32}\left( C^t_L + e_{H_3} \sin^2\theta_W + C^d_L \right) \nonumber \\
 & 
+  \bar s  \gamma^\mu P_R d\, \ln\Lambda^2 g^R_{31} g^{R*}_{32}  \left( C^t_R + e_{H_3} \sin^2\theta_W + C^d_R \right) =0\,,
 \end{align}
where the vanished result arises from $-e_t+e_{H_3}-e_d=0$. 

It was mentioned earlier that the contribution of electromagnetism-like $Z$ coupling to top-quark vanishes at $k^2=0$. In order to verify this result,  we can focus on the effects of $e_f \sin^2\theta_W$  that appear in the $Z$-coupling. Thus, using $e_t = e_{H_3} - e_d$, the renormalized vertex function can be expressed as:
 \begin{align}
 i \Gamma^\mu_{ZR} & \supset  -i \frac{g e_t  \sin^2\theta_W}{(4\pi)^2\cos\theta_W}  \bar s \gamma^\mu \chi^V_{21} d  \nonumber \\
& \times \left[ \frac{1}{2} + \int^1_0 dx (1-2x) \ln(x + y_t (1-x)) - \frac{y_t (1-x)}{x + y_t (1-x)} \right]=0
 \end{align}
 It can be found that the vanished result indeed is similar to that shown in Eqs.~(\ref{eq:gaugeQCD}) and (\ref{eq:gaugeQED}). Hence, the renormalized three-point vertex can be obtained as:
\begin{equation}
i \epsilon^Z_\mu \Gamma^\mu_{ZR}  = -i \frac{g I^t}{ (4\pi)^2 \cos\theta_W } \bar s \slashed{\epsilon}^Z \left( g^L_{31} g^{L*}_{32} I_{Z}(y_t) P_L  - g^R_{31} g^{R*}_{32} I_{Z}(y_t) P_R \right) d\,, \label{eq:L_Z}
\end{equation}
where the resulted vertex function is associated with the top-quark weak isospin $I^t=1/2$, and the loop integral $I_{Z}(y_t)$ is defined as:
 \begin{align}
%I_{ZL}(y) & =  - \frac{1+y}{4(1-y)} + \frac{y(y-2)\ln y }{2(1-y)^2}\,, \nonumber \\
%I_{ZL}(y) & =  - \frac{y}{(1-y)} - \frac{y\ln y }{(1-y)^2}\,, \nonumber \\
%
I_{Z}(y) & = - \frac{y}{1-y} - \frac{y \ln y}{(1-y)^2}\,. \label{eq:I_Z}
\end{align}

\section*{Acknowledgments}

This work was partially supported by the Ministry of Science and Technology of Taiwan,  
under grants MOST-106-2112-M-006-010-MY2 (CHC).

\end{document}